\documentclass[times,english]{smrauth}

\usepackage[T1]{fontenc}
\usepackage[utf8]{inputenc}
\usepackage{array}
\usepackage{booktabs}
\usepackage{textcomp}
\usepackage{graphicx}
\usepackage{enumerate}
\usepackage{babel}
\usepackage[colorlinks,bookmarksopen,bookmarksnumbered,citecolor=red,
urlcolor=red]{hyperref}
\usepackage{cuted}
\usepackage{flushend}
\usepackage{caption}
\usepackage{floatrow}
\usepackage{moreverb}
\usepackage{threeparttable}

\DeclareNewFloatType{example}{name=Example Box}
\DeclareNewFloatType{subexample}{name=Figure}

\newcommand\BibTeX{{\rmfamily B\kern-.05em \textsc{i\kern-.025em b}\kern-.08em
T\kern-.1667em\lower.7ex\hbox{E}\kern-.125emX}}

\def\volumeyear{2013}

\begin{document}

\runningheads{M.~Unterkalmsteiner \itshape{et al.}}{A conceptual framework for
SPI evaluation}

\title{A conceptual framework for SPI evaluation}

\author{
Michael~Unterkalmsteiner\affil{1}\corrauth,
Tony~Gorschek\affil{1},
A.\,K.\,M.~Moinul~Islam\affil{2},
Chow~Kian~Cheng\affil{3}, 
Rahadian~Bayu~Permadi\affil{4} and
Robert~Feldt\affil{1}}

\address{
\affilnum{1}Software Engineering Research Lab, School
of Computing, Blekinge Institute of Technology, SE-371~79 Karlskrona,
Sweden. \break
\affilnum{2}Software Engineering: Process and Measurement Research Group,
Department of Computer Science, University of Kaiserslautern, PO Box 3049,
67653 Kaiserslautern, Germany.\break
\affilnum{3} General Electrics Healthcare, Healthcare IT, Munzinger Straße 5,
79111 Freiburg, Germany.\break
\affilnum{4} Amadeus S.\,A.\,S., Product Marketing and Development, 485 Route
du Pin Montard, Boite Postale 69, 06902 Sophia Antipolis Cedex, France.}

\corraddr{Michael Unterkalmsteiner, Blekinge Institute of Technology, SE-371 79
Karlskrona, Sweden. E-mail: mun@bth.se}

\begin{abstract}
Software Process Improvement (SPI) encompasses the analysis and modification of
the processes within software development, aimed at improving key areas that
contribute to the organizations' goals. The task of evaluating whether the
selected improvement path meets these goals is challenging. Based on the results
of a systematic literature review on SPI measurement and evaluation practices,
we developed a framework (SPI-MEF) that supports the planning and implementation
of SPI evaluations. SPI-MEF guides the practitioner in scoping the evaluation,
determining measures and performing the assessment. SPI-MEF does not assume a
specific approach to process improvement and can be integrated in existing
measurement programs, refocusing the assessment on evaluating the improvement
initiative's outcome. Sixteen industry and academic experts evaluated the
framework's usability and capability to support practitioners, providing
additional insights that were integrated in the application guidelines of the
framework.
\end{abstract}


\maketitle

\section{Introduction}
With the increased importance of software in product
development~\cite{kitchenham_software_1996}, the software engineering discipline
and the study of the involved processes have started to gain more popularity
among researchers and practitioners in
industry~\cite{scacchi_process_2001,sjoberg_future_2007,wirth_brief_2008}.
Software Process Improvement (SPI) encompasses the assessment and improvement of
the processes and practices involved in software
development~\cite{card_research_2004}. SPI involves the understanding of the
software processes as they are used within an organization and suggests areas
for improvements in achieving specific goals such as increasing product quality,
operation efficiency and cost reduction~\cite{oconnor_using_2007}. The SPI
literature provides many case studies of successful companies and descriptions
of their SPI programs~\cite{dyba_instrument_2000}. Examples are presented
by~\cite{canfora_applying_2006,ferreira_roi_2008,goldenson_demonstrating_2003,
hyde_intangible_2004,mohagheghi_empirical_2008,murugappan_blending_2003,
redzic_six_2006,sommerville_empirical_2005,trienekens_product_2001}, and also
covers the recently popular development practices classified as agile or
lean~\cite{achatz_industrial_2003,salo_integrating_2005}. 

Assessing the outcomes of SPI initiatives is as important as their actual
implementation since without a clear understanding of gains or losses, it is
impossible to reason about the performance of an SPI
initiative~\cite{jones_economics_1996}. Measurement in SPI can be of
descriptive, evaluative or predictive nature~\cite{briand_practical_1996}.
Descriptive and predictive measurement is the primary facility to enable the
software process to perform with predictable performance and capability and to
ensure that process artifacts meet their
requirements~\cite{paulish_case_1994,florac_measuring_1999}. Evaluative
measurement aims at providing support for operative
decisions~\cite{briand_practical_1996}. In this paper we focus on the evaluative
nature of measurement, targeted at assessing the impact of SPI initiatives. 

The success of improvement initiatives also means different things to different
people~\cite{freeman_measuring_1992}. Hence, various stakeholders’ points of
view have to be taken into consideration when assessing the outcome on an SPI
program~\cite{abrahamsson_measuring_2000}. The causal relationship between the
improvement initiative and its effect is complex, and it is hard to determine
whether the effect being measured is stemming exclusively from the improvement
initiative~\cite{dror_process_2007}. 

The lack of guidelines for conducting evaluative SPI measurements have raised
the challenge to develop and implement effective performance measurement
programs for SPI~\cite{iversen_problems_2006}. Since the evaluation of the
outcome of an SPI initiative is complex but also crucial to the organization,
there is a need for a measurement and evaluation framework which guides SPI
practitioners in their work, helping in preserving effort and cost, and enabling
return on investment to be ascertained. Such a framework should promote an
evaluation which considers the improvement from different views, increase the
visibility, and consequentially facilitate the assessment of the achieved
benefits. 

The challenges in process improvement evaluation are diverse, ranging from
defining an appropriate measurement scope, eliciting the required metrics, to
the consideration of confounding factors in the
evaluation~\cite{unterkalmsteiner_evaluation_2012}. This paper proposes a
conceptual framework that aims to address these challenges, offering a
structured approach. The framework was derived from an extensive systematic
literature review~\cite{unterkalmsteiner_evaluation_2012} which collected best
practices in the field. Subsequently, we followed the technology transfer model
proposed by Gorschek et al.~\cite{gorschek_model_2006} to statically validate 
the framework by experts from both academia and industry.

The remainder of this paper is organized as follows. Related work is discussed
in Section~\ref{sec:Related-work}. In
Section~\ref{sec:Challenges-in-measuring-and-evaluating-SPI-initiatives} we
present four major challenges, basing their formulation on the results of a
systematic literature review on SPI measurement and
evaluation~\cite{unterkalmsteiner_evaluation_2012}. With the aim to address
those challenges, we developed the Software Process Improvement Measurement and
Evaluation Framework (SPI-MEF). Section~\ref{sec:SPI-MEF} describes the
framework, and an example scenario is provided
in~\cite{unterkalmsteiner_extended_2011}. The usefulness and usability of
SPI-MEF was validated through the help of 9 research experts and 7 industry
practitioners as described in Section~\ref{sec:Validation}. The results
and the refinements applied to SPI-MEF are discussed in
Section~\ref{sub:Results}. Threats to validity are discussed in 
Section~\ref{sub:Threats-to-validity}. Finally, conclusions and
motivations for future work are given in Section~\ref{sec:Conclusion}.

\section{Related work}\label{sec:Related-work}
In this section, we briefly review previous work relevant to the measurement and
evaluation framework proposed in this paper. 

Software process appraisal methods, e.g. SCAMPI~\cite{_standard_2011}, or
guides to process assessment, e.g. ISO/IEC 15504 (Part
4)~\cite{_iso/iec_1998}, evaluate whether an organization conforms to a certain
industry standard. The assessment identifies areas for
improvement~\cite{ares_more_2000} and can steer the implementation of process
improvements~\cite{ekdahl_experience_2006}. Such assessments provide a benchmark
against a set of goals, do however not evaluate the actual impact of process
changes.

SPI research into measurement programs has developed and suggested several
metrics~\cite{fenton_metrics_2001}. For example, the \emph{ami} (Assess,
analyze, Metricate, Improve) approach integrates an analytic, bottom-up with a
benchmarking, top-down approach to process
improvement~\cite{pulford_quantitative_1995,
kuntzmann-combelles_quantitative_1995}. The rationale for the expected synergy
is that top-down approaches, such as the Capability Maturity Model
(CMM)~\cite{paulk_capability_1993}, do not consider the specific business goals
of a company~\cite{debou_linking_2000}. Hence, the proposed goal-oriented
measurement in ami, based on the GQM
paradigm~\cite{basili_methodology_1984,basili_goal_1994}, serves to analyze the
identified improvement opportunities more in depth and to monitor the
implemented changes, assuring that the followed best practices lead to the
achievement of the targeted business goals. Similarly, the GQM+Strategies
approach aims at linking business strategies with measurement
goals~\cite{basili_gqm^+_2007}, since CMMI~\cite{_capability_2010} does not
provide an explicit link from the improvement to business
value~\cite{basili_linking_2010}. 

Inspired by the ami approach, Park et al.~\cite{park_goal-driven_1996} developed
the GQ(I)M method. Extending the GQM paradigm, GQ(I)M introduces the notion of
\emph{indicators}, which reflect the idea of asking ``What do I want to know?''
as opposed to the question ``What do I want to measure?''. Indicators are
therefore representations of measurement data, backed by one or more metrics,
and support with a clear definition of their construction the decision making
process~\cite{goethert_applications_2004}.

The integration of process assessment, modeling and measurement is the goal of
the product-focused improvement approach (PROFES)~\cite{jarvinen_profes_2000}.
As opposed to ami and GQ(I)M, in which \emph{company specific} business goals
define the measurement strategy, PROFES promotes continuous assessment against
\emph{reference models} such as CMM or ISO/IEC 15504, supported by measurements
derived by GQM~\cite{jarvinen_establishing_1999,jarvinen_integrating_1999}. The
expected benefits are higher visibility of process changes and therefore better
control on the improvement process and lower assessment costs, as the time
needed for data collection is reduced~\cite{jarvinen_establishing_1999}.

An alternative to the ubiquitous GQM paradigm is the Practical Software and
System Measurement (PSM) approach~\cite{mcgarry_practical_2001}, that 
influenced and was influenced~\cite{card_status_2003} by the in parallel 
developed international standard for Software Process Measurement, ISO/IEC
15939~\cite{international_organization_for_standardization_iso/iec_2002}. In 
contrast to the more general, goal-oriented GQM, PSM is designed to establish a 
measurement process for project evaluation, following the Plan-Do-Check-Act 
cycle~\cite{card_status_2003}. Measurement in PSM has the purpose to satisfy 
the project manager's information needs which stem from a) the achievement of 
project success, and b) obstacles or issues related to achieving 
success~\cite{bailey_psm_2003}.

We reviewed the literature on SPI evaluations conducted in
industry~\cite{unterkalmsteiner_evaluation_2012}, identifying current practices
that also were built upon, or were inspired by the approaches discussed in this
section. The analysis of these practices lead to the definition of several
challenges related to the evaluation of SPI initiatives which are discussed in
further detail in
Section~\ref{sec:Challenges-in-measuring-and-evaluating-SPI-initiatives}.

\section{Challenges in measuring and evaluating SPI
initiatives}\label{sec:Challenges-in-measuring-and-evaluating-SPI-initiatives}
Obstacles and issues in implementing measurement programs in general were
previously identified by Herbsleb and Grinter~\cite{herbsleb_conceptual_1998}
(difficult communication across organizational boundaries, rigidity of data
collection mechanisms, non-transparent data usage), Berry and 
Ross~\cite{berry_instrument_2000} (the complexity of combining sociological and 
technological aspects in a measurement program), Kasunic et 
al.~\cite{kasunic_can_2008} (poor data quality), and Umarji and
Seaman~\cite{umarji_gauging_2009} (different perceptions of metrics between
developers and managers).

Since the focus of these issues is predominantly on the implementation of
measurement programs, which is a critical but not the sole aspect of SPI
evaluation, we devised four fundamental challenges in measuring and evaluating
SPI initiatives. The formulation of these challenges bases upon the findings
from a systematic literature review on measurement and evaluation of
SPI~\cite{unterkalmsteiner_evaluation_2012}.

Sections~\ref{sub:Challenge-I} to~\ref{sub:Challenge-IV} characterize the
identified challenges and explain how they are addressed by the six concepts
presented in SPI-MEF.

\begin{figure}
\begin{centering}
\includegraphics[scale=0.5]{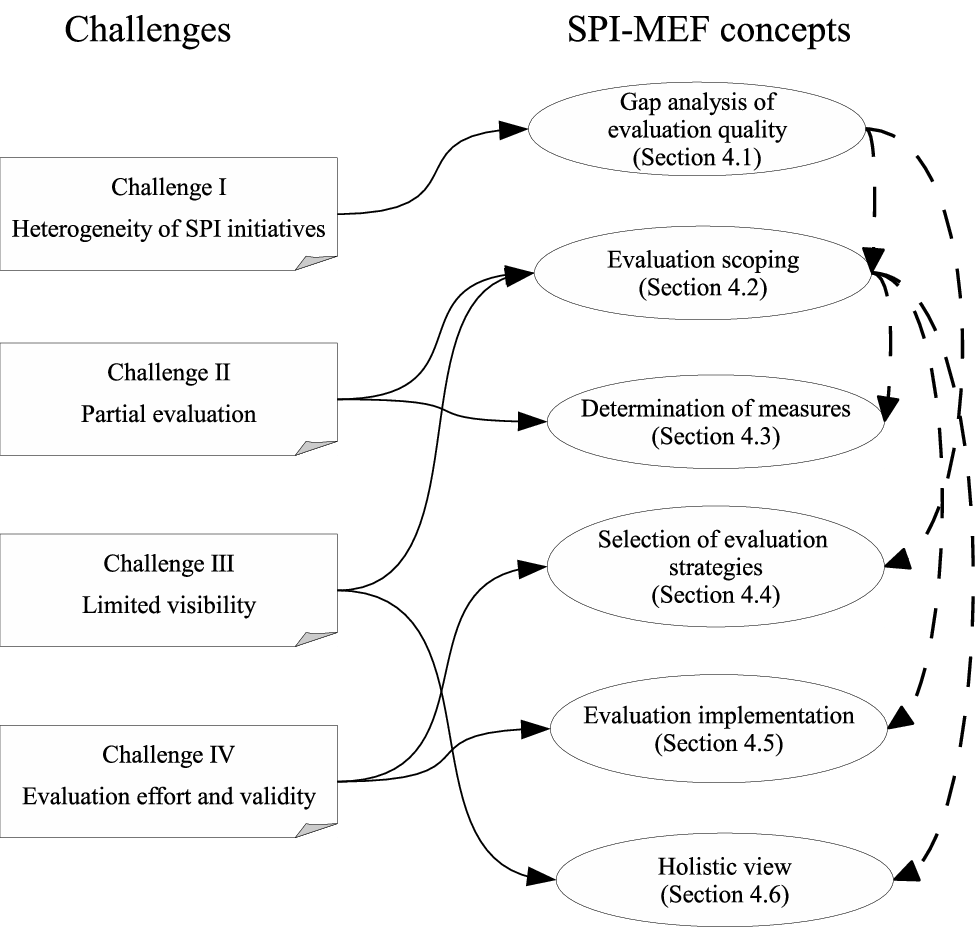}
\par\end{centering}
\caption{Conceptual map of the framework\label{fig:Conceptual-map}}
\end{figure}

\subsection{Challenge I - Heterogeneity of SPI
initiatives}\label{sub:Challenge-I}
The spectrum of SPI initiatives ranges from the application of tools for
improving specific development processes, to the implementation of
organization-wide programs to increase the software development capability as a
whole~\cite{unterkalmsteiner_evaluation_2012}. As a consequence of this variety
and diversity in scope and complexity of SPI initiatives, we designed SPI-MEF as
a set of interrelated concepts, each one addressing one or more challenges.
Figure~\ref{fig:Conceptual-map} summarizes the relationships between challenges
and concepts.

In each concept we provide a set of practices which can be used to fulfill the
goals of the concept and hence addressing the challenge. These practices however
may need to be adapted and scaled to the specific context in which the framework
is used. Hence, the first concept is termed \emph{Gap analysis of evaluation
quality}. It provides means to assess the current and to define the aspired
evaluation quality, enabling the customization and scaling of the framework to
different types of SPI initiatives.

\subsection{Challenge II - Partial evaluation}
The outcome of SPI initiatives is predominately assessed by evaluating measures
which are collected at the project
level~\cite{unterkalmsteiner_evaluation_2012,gorschek_requirements_2008}. As a
consequence, the improvement can be evaluated only partially, neglecting effects
which are visible only outside individual projects. Such evaluations can
therefore lead to sub-optimizations of the
process~\cite{herbsleb_conceptual_1998}. By focusing on the measurement of a
single attribute, e.g. effectiveness of the code review process, other
attributes might inadvertently change, e.g. time-to-market of a product. 

To address this challenge, we propose the concept of \emph{Evaluation scoping}
which provides means to determine the extent of the improvement evaluation. This
concept provides the answer to the question: where to measure? Complementary we
also propose the concept of \emph{Determination of measures} which aims at
providing the answer to the question: what to measure?

\subsection{Challenge III - Limited visibility}
This challenge is a consequence of the previous one since a partial evaluation
implies that the gathered information is targeted to a specific audience which
may not cover all important stakeholders of an SPI initiative. This means that
information requirements may not be satisfied, and that the actual achievements
of the SPI initiative may not be visible to some stakeholder as the measurement
scope~\cite{unterkalmsteiner_evaluation_2012} is not adequately determined.

\emph{Evaluation scoping} aims to address this issue by providing a structured
approach to identify the relevant stakeholders and to provide them with the
information they need. The concept of \emph{Holistic view}, on the other hand,
provides a way to collect and present the gathered information, supporting a
multi-faceted view on the improvement initiative.

\subsection{Challenge IV - Evaluation effort and
validity}\label{sub:Challenge-IV}
Due to the vast diversity of SPI initiatives (see Challenge I), it is not
surprising that the evaluation strategies vary. The evaluation and analysis
techniques are customized to the specific settings were the initiatives are
embedded~\cite{unterkalmsteiner_evaluation_2012}. Since there exist no formal
guidelines for implementing an SPI evaluation~\cite{iversen_problems_2006}, one
can assume that the design and development of the evaluation strategies require
a considerable amount of effort. Furthermore, confounding factors are seldom
taken into account in the industrial practice of improvement
evaluation~\cite{unterkalmsteiner_evaluation_2012}. This can be a major threat
to the evaluation validity since the predominant practice of improvement
evaluation is based on pre-post
comparison~\cite{unterkalmsteiner_evaluation_2012}. 

To address this challenge, the concept \emph{Selection of evaluation strategies}
provides support in identifying and implementing adequate means for SPI
evaluation. In addition, the concept \emph{Evaluation implementation} discusses
timing factors that should be considered and provides support for conducting the
evaluation itself.

\subsection{Summary}
Figure~\ref{fig:Conceptual-map} shows a conceptual map, indicating how Challenge
I to IV are addressed by the SPI-MEF concepts, and how the concepts are
related to each other. \emph{Gap analysis of evaluation quality}, whose
intent is to tune the overall measurement and evaluation approach, is directly
connected to \emph{Evaluation scoping} and \emph{Selection of evaluation
}strategies, and indirectly to \emph{Determination of measures}. Similarly,
\emph{Evaluation scoping}, whose intent is to define where to measure and who
will see the results, influences the \emph{Evaluation implementation}, and the
\emph{Holistic view} concepts. In Section~\ref{sec:SPI-MEF} we describe all
SPI-MEF concepts in detail, and present practices used to realize the concept in
practice.

\section{SPI-MEF}\label{sec:SPI-MEF}
SPI-MEF was developed based on an extensive study of SPI research and industry
case studies~\cite{unterkalmsteiner_evaluation_2012}, mapping best practices,
but also gaps in knowledge. The reviewed primary studies provided an excellent
source of practices applied successfully in industry, but even more importantly,
allowed us to extract generic guidelines to support the evaluation of SPI
initiatives.

This section presents these guidelines, complemented with 9 interconnected,
although fictitious, examples (starting with Example Box~\ref{exa:Example-1}),
which in essence constitute SPI-MEF. As the relationships shown in
Figure~\ref{fig:Conceptual-map} suggest, there exist dependencies between the
concepts. They are however ordered in a way, reflected in the structure of the
guidelines and the examples, in which one would typically conduct the planning
for the evaluation. The extended scenario
provided in~\cite{unterkalmsteiner_extended_2011} shows how the framework is
applied in an iterative manner, following a phased approach.

\begin{example}[b]
\caption{Introduction}\label{exa:Example-1}
The scenario is embedded within the context of a medium-sized software
development organization called ALPHA. ALPHA has 70 full-time software
professionals developing off-the-shelf software applications for various
industries. Software development in the company follows an iterative spiral
life-cycle model and a typical software product is released after 6 months.

ALPHA develops products on top of a software platform. ProductA has just been
released and incremental feature releases are scheduled bi-yearly. Due to
quality assurance issues at the end of the development life-cycle of ProductA,
it was decided to introduce code inspections in the projects for the feature
releases. As planning for ProductB has commenced and new development staff was
employed, code inspections were also deemed to be an effective means to
introduce new developers to the software platform.
\end{example}

\subsection{Gap analysis of evaluation
quality}\label{sub:Gap-analysis-of-evaluation-quality}
This concept, addressing \emph{Challenge I - Heterogeneity of SPI initiatives},
reflects the need to capture the context in which SPI initiatives are
implemented. The particular context characteristics steer further decisions,
e.g. in Evaluation scoping (Section~\ref{sub:Evaluation-scoping}) and Selection
of evaluation strategies (Section~\ref{sub:Selection-of-evaluation-strategies}).
The basic information that should be recorded is (see also Example
Box~\ref{exa:Example-2}): 
\begin{enumerate}[(a)]
\item a description of the initiative and its purpose
\item concrete improvement goals
\item the affected process areas
\item the target entities of the initiative, i.e. specific projects,
products or departments, and 
\item a tentative schedule for the implementation.
\end{enumerate}
Furthermore, the organization's capability to implement a measurement program
and conduct an evaluation needs to be assessed by a gap analysis. In general,
gap analysis uses two sets of information: the current status and the aspired
status~\cite{stalhane_root_2004}. The current status of measurement capability
can be determined by following one of the maturity assessment approaches
presented by Daskalantonakis et al.~\cite{daskalantonakis_method_1990}, Comer
and Chard~\cite{comer_measurement_1993}, Niessing and
Vliet~\cite{niessink_towards_1998}, and Diaz-Ley et
al.~\cite{diaz-ley_mis-pyme_2008}.
Then the aspired measurement and evaluation quality has to be determined. The
subsequently identified gap then shows what refinements are needed in the
measurement program. 

\begin{figure}
\begin{centering}
\includegraphics[scale=0.4]{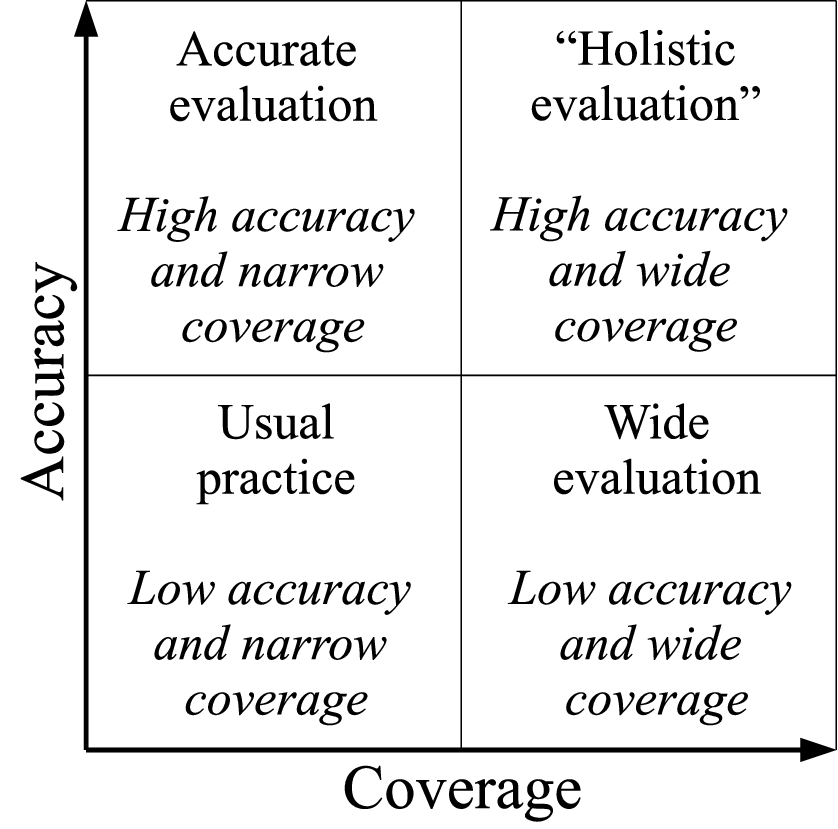}
\par\end{centering}
\caption{Opportunity matrix\label{fig:Opportunity-matrix}}
\end{figure}

\begin{example}[b]
\caption{Initiative context}\label{exa:Example-2}
The current software development processes in ALPHA offers several opportunities
for improvement. A process team has been set up to investigate the process areas
that need attention, and they proposed to introduce code inspections. The
initiative's context is documented below: \medskip{}

\begin{center}
\begin{tabular}{>{\raggedright}m{1in}>{\raggedright}m{4in}}
\toprule
Type of SPI initiative & Practice (Code inspections)\tabularnewline
\midrule
Description & Code inspection is a systematic process in reviewing one
developer’s work product in the coding phase. The code inspection follows a
structured process that involves the planning, preparation, inspection meeting,
rework and follow-up. The code written by the developer will be inspected by two
or more peers, usually more experienced or senior developers, in the project
team.\tabularnewline
\midrule
Improvement goal(s) & Improve product quality\tabularnewline
\midrule
Process areas / phases & Coding phase\tabularnewline
\midrule
Target entities & All development projects\tabularnewline
\midrule
Implementation schedule & Phase I: Pilot projects

Phase II: All projects in the organization\tabularnewline
\bottomrule
\end{tabular}
\par\end{center}
\end{example}

SPI-MEF proposes to use a “2x2” matrix (Figure~\ref{fig:Opportunity-matrix}) to
support the decision process, using the context information and current
measurement capability as input. The evaluation quality in SPI-MEF is defined by
two dimensions: accuracy and coverage. Accuracy can be improved by considering
primary and complementary measures
(Section~\ref{sub:Primary-and-complementary-indicators}), selecting the
appropriate evaluation strategy and by taking confounding factors into account
(Section~\ref{sub:Selection-of-evaluation-strategies}). Coverage is determined
by to what extent measurement levels and viewpoints
(Section~\ref{sub:Evaluation-scoping}) are included in the evaluation. It is
possible to address both dimensions simultaneously, but cost and effort
constraints may prohibit such a strategy. The “2x2” matrix has been proven to be
an excellent tool to address such decision dilemmas~\cite{lowy_power_2004}. The
roles involved in the discussion about the long-term strategy should include
management which provides the funding for the improvement initiative and
measurement program experts. The Evaluation Opportunity Matrix
(Figure~\ref{fig:Opportunity-matrix} and Example Box~\ref{exa:Example-3}) serves
as a starting point for the discussion. 

Aside from accuracy and coverage another important aspect to consider is the
cost of the evaluation. Cost is denoted as a function of the quality and scope
of the evaluation. Therefore, the resources an organization is willing to
invest have to be taken into consideration when choosing the desired path to
improve the quality of evaluation. The cost of evaluation can arise from the
amount of metrics defined and collected, the resources needed to manage the
metrics, the number of people involved in the metric collection and evaluation,
etc. Although achieving high accuracy and coverage in the context of SPI-MEF
seems to inherently require more metrics, the potential reuse of metrics should
be considered when evaluating the cost for evaluation implementation. During the
decision process of which strategy to follow, it is advisable to involve
personnel who have expertise in implementing a measurement program and can
estimate the cost of metric collection, and management, which sponsors the
improvement program. 

\begin{example}[h]
\caption{Evaluation opportunity matrix}\label{exa:Example-3}
\begin{minipage}{0.5\linewidth}
The Software Engineering Process Group (SEPG) at ALPHA met with the upper level
management and employees who are currently in charge of the measurement program.
To increase accuracy in the short term, plans are to consider primary and
complementary measures and to control the major confounding factors by
establishing a baseline from the appropriate historical data. Considering the
process improvement budget and the implementation schedule, the company’s
short-term goal is to focus on Process and Project level accuracy first and then
address coverage by including the Product measurement level.
\par
\end{minipage}
\hspace{0.4in}
\begin{minipage}{0.5\linewidth}
\includegraphics[scale=0.3]{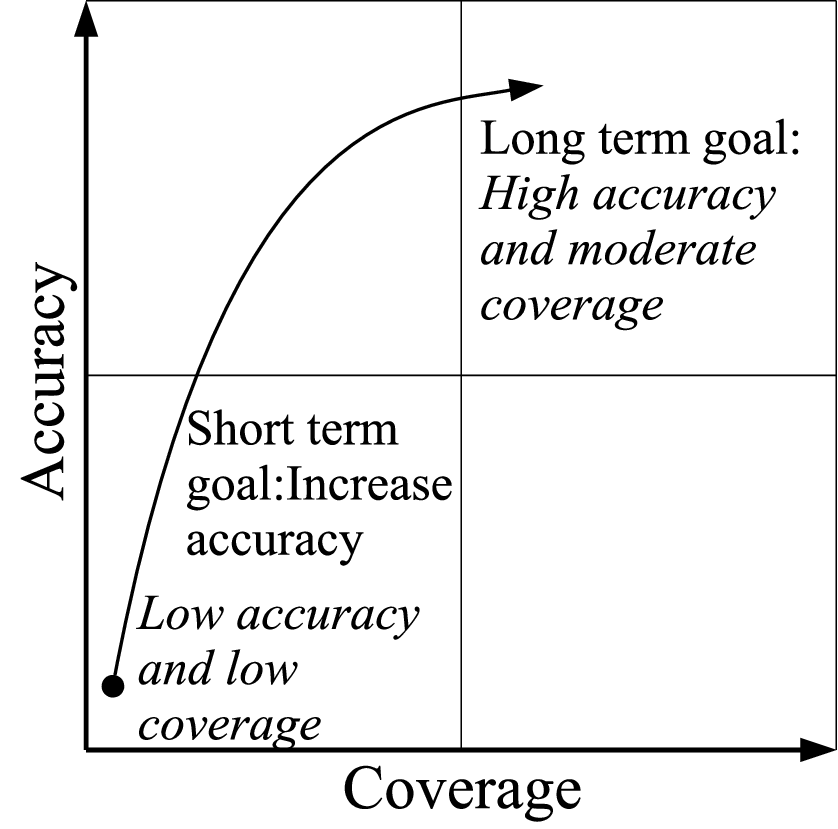}
\end{minipage}
\end{example}

\subsection{Evaluation scoping}\label{sub:Evaluation-scoping}
The evaluation scope is determined using two dimensions. The Measurement Levels
(MLs) represent the spectrum of measurable entities which can potentially be
assessed in the evaluation. Identifying the Measurement Levels for evaluation
counteracts \emph{Challenge II - Partial evaluation }as it leads the
practitioner to consciously define the coverage of the evaluation.
Section~\ref{sub:Measurement-Levels} explains the Measurement Levels in more
detail.

The second dimension, Evaluation Viewpoints (EVs), represent the stakeholders
and their information needs in relation to the evaluation of the improvement.
Defining Evaluation Viewpoints counteracts \emph{Challenge III - Limited
visibility} as it clarifies the stakeholders' data requirements to evaluate the
improvement initiative. Section~\ref{sub:Evaluation-viewpoints} discusses the
Evaluation Viewpoints in more detail.

\subsubsection{Measurement Levels.}\label{sub:Measurement-Levels}
The Measurement Levels (MLs) represent the spectrum of entities which are
affected by SPI initiatives and need to be measured in order to achieve a
holistic evaluation of the SPI initiative’s outcome. The MLs, Process,
Project, Product, Organization and External are inspired by the levels of
dependent variables proposed by Gorschek and
Davis~\cite{gorschek_requirements_2008}.

On the \emph{Process} level, the efficiency and effectiveness of the implemented
process improvement initiative can be assessed. For example, if the process
change consists of involving testers in requirements reviews, it can be measured
how many faults are identified compared to the previous instance of the process.
A measurable gain at this level is however not sufficient to assert that the
improvement goal (e.g. improved product quality) has been reached. Furthermore,
the improvement of one process may produce side-effects on other processes, or,
more generally, affect the output of the process, which is not measurable at
this level. 

The measurement at the \emph{Project} level is mainly concerned with project
control by monitoring budget, schedule and resources. A projects' success or
failure is often evaluated by determining the discrepancy between estimated and
actual values. Additionally, it is possible to measure the effects of newly
introduced or modified processes by assessing the work products created during
the project. For example, if the requirements review process leads to fewer
specification changes during the project life-cycle. Adherence to project
estimates can indicate process improvement but can also be misleading when
considered in isolation as product quality is not assessed.
Linberg~\cite{linberg_software_1999} reports on a case study in which a project
faced severe schedule and budget overruns. From the management's point of view,
the project was perceived as a failure, whereas the product, once shipped, was
highly successful. Thus, considering the project \emph{and} the product
perspective in an improvement evaluation is important.

Increasing product quality is often the major improvement goal when establishing
an SPI initiative. Measurement at the \emph{Product} Level assesses both
internal quality attributes which are mostly visible to software developers, and
external quality attributes which are observed by the user of the product.
Besides increased quality, process improvement may also target a reduction in
cost and time-to-market of the product. Continuing on the previous example with
requirements reviews, the involvement of testers could lead to a delay in other
projects, to which they were originally assigned. The project with the improved
review process and tester involvement could be completed earlier due to less
rework; other projects however could be delayed due to the deduction of
resources. It is therefore necessary to control and assess all aspects of the
improvement goals and to take them into consideration when evaluating the
initiative's success. 

The short- and mid-term effects of an SPI initiative can be assessed in the
Process, Project and Product level, but the long-term effects will prevail and
only be visible at the \emph{Organization} level. An SPI initiative has to meet
the business goals of a company and has to be aligned with its vision.
Therefore, it is necessary to assess the improvements impact on the
organization's business strategy, economy and culture. The example with the
involvement of testers in requirements reviews shows that a reorganization of
the development process may be required in order to avoid resource and
scheduling issues. Hence, the measurement and evaluation of the performance of
the SPI initiative at the Organization level is of importance for the design and
implementation of forthcoming process improvements.

The previously mentioned MLs are focused on the measurement and evaluation of
the SPI initiative within the company and neglect that the effect of the
improvement may also transcend the organizational border to the exterior world.
The \emph{External} level is influenced by the produced goods but also by the
organization itself, e.g. through supplier dependencies. For example, the
aforementioned improvement of the development process can also affect suppliers
as they may need to interact differently with their client. Measurement at the
External level assesses positive and negative externalities which should be
taken into consideration when evaluating the success of an SPI initiative.

\subsubsection{Effect traceability in Measurement
Levels.}\label{sec:Effect-Traceability}
Effect traceability in an issue inherent in the Measurement Levels that was also
brought up by Gorschek and Davis~\cite{gorschek_requirements_2008}. The
traceability between the action and its empirically assessable effects
diminishes with increasing distance from the process change. 

Due to temporal distance there is an increasing latency by which the effect of
process improvement is measurable at the different levels. That is, the effect
of the treatment will reach the process itself first, then the projects in which
the process is applied and eventually the products emerging from the various
projects. 
Furthermore, the ability to isolate the effect on a particular improvement
decreases. Each level is an aggregation of one or more entities of the previous
level, e.g. the External level includes, besides other things, a set of
organizations, which in turn, include, besides other things,  a set of products.
These "other things" can be seen as external variables and are defined in  this
context as confounding factors. Those may aggravate an accurate evaluation
because they  hide or amplify the effects of the improvement initiative. 

In order to counteract these effect traceability issues we propose
Evaluation Viewpoints as the second dimension for evaluation scoping.

\subsubsection{Evaluation Viewpoints.}\label{sub:Evaluation-viewpoints}
According to Zahran~\cite{zahran_software_1998}, a software process improvement
initiative has to be backed up by both organizational and management
infrastructure, as well as a process technical infrastructure. The
organizational and management infrastructure defines the stakeholders which are
usually involved in the improvement initiative, such as executive sponsors, a
steering committee, a software engineering process group (SEPG), and software
process improvement teams. Besides the viewpoints represented by the previously
mentioned SPI stakeholders, the evaluation should also consider the viewpoints
from top- and middle- management, product and project management, and software
developers which are not directly in charge of the improvement
initiative. Daskalantonakis~\cite{daskalantonakis_practical_1992} identified six
target audiences for the evaluation and use of metrics in software
organizations: Software users, Senior Managers, Software Managers, Software
Engineers, and Software Process Engineers and Software Quality Assurance.
Similarly, Ebert~\cite{ebert_technical_1999} identified four roles with
individual goals related to the improvement: practitioners, project managers,
department head, and corporate executives.

\begin{example}[b]
\caption{Evaluation scoping}\label{exa:Example-4}
In Example Box~\ref{exa:Example-3}, the SEPG decided to evaluate the improvement
initiative, code inspections, on the Process, Project and Product levels. The
SEPG defines the Evaluation Viewpoints (EVs) for the respective levels. For
illustrative purposes, the table in this example contains also the Organization
and External levels. The first column denotes the Measurement Levels (MLs),
i.e. the entities that are affected by the SPI initiative, whereas the remaining
columns denote the Evaluation Viewpoints (EVs), i.e. the stakeholders that have
an interest in the evaluation of the SPI initiative. The table is read, for
example, as follows: ``The development team is interested in evaluating the
impact of the SPI initiative on the Process ML from the Implementer EV.''
Note that a specific role can have different interests when
evaluating an improvement initiative and therefore represent more than one
viewpoint. By looking at the table, the ``Product Manager'' role subsumes both
the Coordinator viewpoint at the Product level and the Implementer viewpoint at
the Organization level.

\begin{center}
\begin{tabular}{c>{\centering}m{1.2in}>{\centering}m{1.2in}>{\centering}m{1.2in}
}
\toprule 
 & \multicolumn{3}{c}{\textbf{EVs}}\tabularnewline
\textbf{MLs} & \multicolumn{1}{c}{\emph{Implementer}} & \emph{Coordinator} &
\emph{Sponsor}\tabularnewline
\midrule 
\emph{Process} & Development team  & SEPG  & SPI steering
committee\tabularnewline
\midrule 
\emph{Project} & Development team  & Project manager / SEPG & SPI steering
committee / Head of department \tabularnewline
\midrule 
\emph{Product} & Development team / Project managers & Product manager & Head of
department\tabularnewline
\midrule 
\emph{Organization} & Product manager  & Board of directors  & Company
shareholders / Customers \tabularnewline
\midrule 
\emph{External} & Product department & Board of directors  & Product user /
Regulator \tabularnewline
\bottomrule
\end{tabular}
\end{center}
The rationale for the mapping in the table is given by considering both the
characteristics of MLs and EVs.
The Implementer viewpoint requires feedback in a short- and midterm time-frame
of the improvement. The development team and project managers, who are
responsible to put the code reviews into practice, represent therefore the
Implementer viewpoint. 
The coordination and control of the improvement activities is the Coordinator's
viewpoint concern. In the case of ALPHA, the SEPG and the Project/Product
manager need to know if the initiative was efficiently implemented and can be
expanded to all projects within the company.
The Sponsor viewpoint, on the other hand, needs confirmation that the
improvement benefits the organization in the mid- to long-term. The SPI steering
committee is interested in identifying conflicts / inefficiencies in the changed
process, whereas the head of department needs to assess the financial payoff.
\end{example}

The specific roles encountered in an organization and in an SPI initiative are
highly dependent on the structure of the organization and the extent of the
process improvement initiative. Hence, we generalize the potential stakeholders
into three Evaluation Viewpoints (EVs): Implementer, Coordinator and Sponsor.
The three EVs reflect the different angles from which the process improvement is
perceived and, more importantly, which aspects of the improvement matter to whom
when conducting the evaluation. The definition of different viewpoints also
supports the idea of increasing the visibility of the process improvement, which
is, presenting the information to the appropriate stakeholders and alleviate the
decision-making process (see Example Box~\ref{exa:Example-4}). It is important
to point out that for a holistic evaluation of the improvement initiative it is
necessary to consider and account for all viewpoints and the respective
evaluation results without isolating single
aspects~\cite{abrahamsson_measuring_2000}. 

The \emph{Implementer} viewpoint represents all the roles which are dedicated to
put the software development in general, and the process improvement in
particular, into practice. The evaluation from this viewpoint is needed to make
the effects of changes in behavior visible to the enactors of the process
improvement. The rationale behind this argument is that a feedback loop on the
effects of the improvement fosters the sustainment of process change.
Additionally, if the Implementer is well informed about the improvement and is
conscious of its effects, he can serve as an accurate data source for the
evaluation of the improvement~\cite{parkinson_practitioner-based_2010}, as well
as be an active contributor to the
improvement~\cite{gorschek_packaging_2004,ivarsson_tool_2012}. 

The \emph{Coordinator} viewpoint comprises the roles which generally participate
in software development and in a software process improvement initiative as
coordination and control entities. Their areas of responsibility include
managing and leading the Implementers, and to steer and promote the process
improvement through strategic (higher level, global), and tactical (lower level,
local), decisions. The interests in evaluating the improvement initiative from
this viewpoint are several, but in general they boil down to two aspects: (a) to
assess if the improvement goals have been achieved, and use the output of the
evaluation to drive and guide further improvement activities, and (b) to provide
feedback to superiors. 

The \emph{Sponsor} viewpoint represents those roles which fund and motivate the
improvement initiative, and in parallel, those who are interested in evaluating
the improvement according to its costs and benefits. The motivating roles' focus
is towards the evaluation of the improvement process itself in order to assess
if it delivered the anticipated benefits. This includes, for example, the SPI
steering committee and/or the head of department. On the other hand, the
evaluating roles' focus may be less on the improvement process itself and rather
on the results which are visible in the environment in which the process change
is embedded. This includes, for example, higher level management financing the
effort, but also company-external entities like shareholders, customers or
regulatory stakeholder. In either case, the evaluation needs to be able to
confirm the long-term effects of the process improvement.

\subsection{Determination of measures}\label{sub:Determination-of-measures}
In order to perform an accurate evaluation, measurements need to provide the
required data. The question is how to elicit the set of sufficient measurements
that allow an evaluation to express reliably if an improvement goal has been
reached or not. The common approach is to derive the required metrics
from the improvement goal. For example, if a reduction in cost is targeted, one
could measure and evaluate if the expended resources in a project which
implements the process improvement were reduced as compared to a previous,
similar, project. 

There are two problems with this approach. 
\begin{enumerate}
\item Not all the benefits in terms of cost reduction may be visible at the
project level, that is, assessing on this level alone would only show a subset
of the achieved benefits.\label{enu:Not-all-the}
\item If the expenditure of resources in a project is the only assessed
dependent variable, it is not possible to evaluate if the improvement did
provoke any side-effects. In particular, detrimental influences that are visible
only with some delay, and on different Measurement Levels, would not be
accounted for in an evaluation based on pure Project level measurements.
Further, using only a single metric as an achievement indicator could raise
validity concerns in the subsequent evaluations~\cite{ramil_defining_2000}. One
reason for this could be data collection issues, e.g. incomplete data-sets or
incorrectly compiled data forms.\label{enu:If-the-expenditure} 
\end{enumerate}
To address problem~\ref{enu:Not-all-the}, one has to reason about selecting the
appropriate target audience for the evaluation and then deriving the necessary
measurements. This evaluation scoping was discussed in
Section~\ref{sub:Evaluation-scoping} were we introduced Measurement Levels and
Evaluation Viewpoints as scoping instruments. 

\begin{figure}
\begin{centering}
\includegraphics[width=30pc]{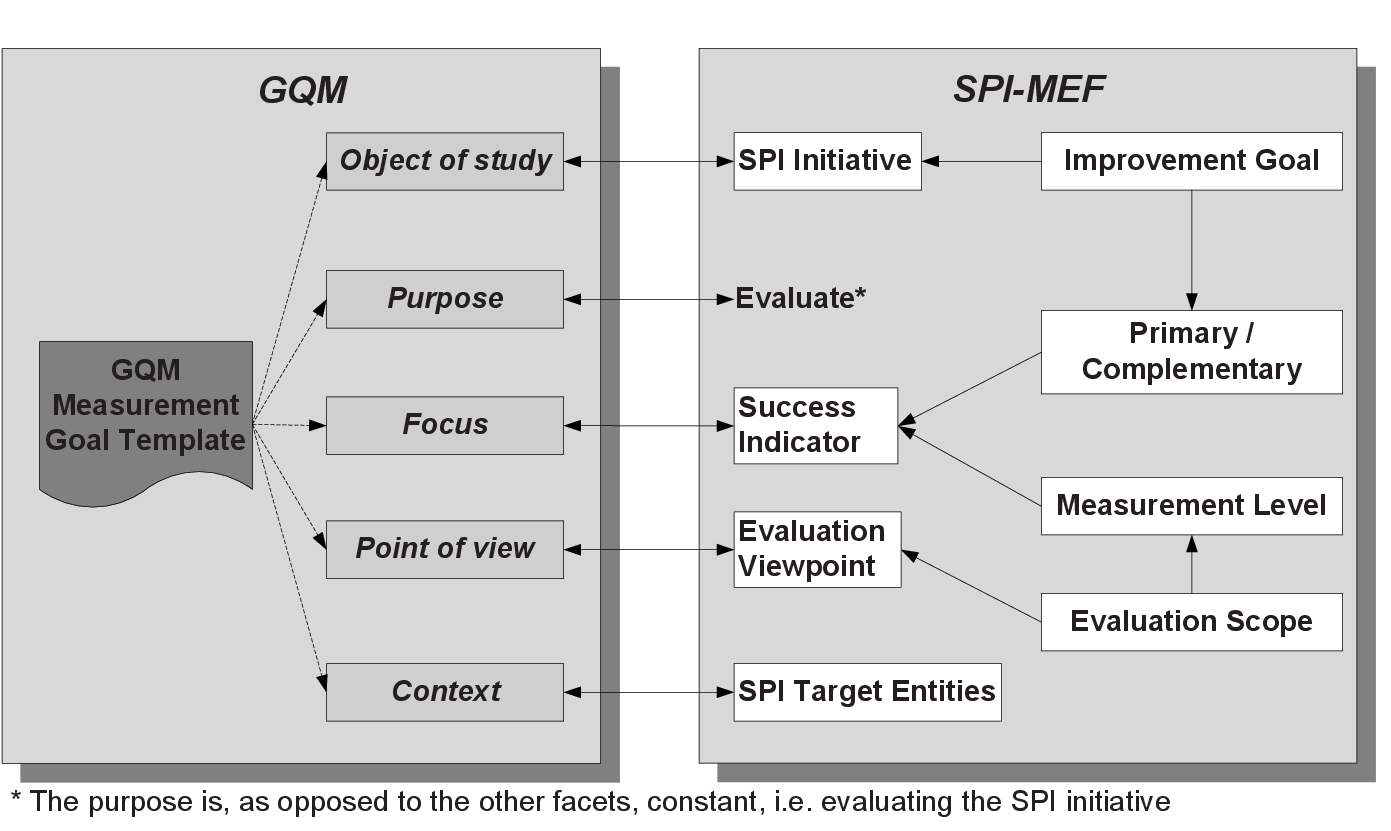}
\par\end{centering}
\caption{SPI-MEF interface with GQM\label{fig:SPI-MEF-interface-with-QGM}}
\end{figure}

To address problem~\ref{enu:If-the-expenditure} we propose a method that builds
upon the Goal-Question-Metric (GQM)
paradigm~\cite{basili_methodology_1984,basili_goal_1994}. GQM is a systematic
way to tailor and integrate an organizations' objectives into measurement goals
and refine them into measurable values. It provides a template for defining
measurement goals and guidelines for top-down refinement of measurement goals
into questions and then into metrics, and a bottom-up analysis and
interpretation of the collected data~\cite{basili_goal_1994}. SPI-MEF provides
an interface with the conceptual level of GQM to define the appropriate
measurement goals for improvement evaluation as illustrated in
Figure~\ref{fig:SPI-MEF-interface-with-QGM}. 

We tailor the GQM approach to the context of SPI measurement and evaluation. The
rationale for interfacing the GQM facets with SPI-MEF, illustrated also in
Example Box~\ref{exa:Example-5}, is as follows. The ``Object of study'' facet
corresponds to the implemented SPI initiative whereas the ``Purpose'' is to
evaluate the impact of the change. The ``Focus'' corresponds to the notion of
success indicator which is determined by the Measurement Level and the
consideration of primary and complementary indicators.
Table~\ref{tab:Mesurement-Levels-and-success-indicators} lists success
indicators that are commonly encountered in SPI initiatives.
Section~\ref{sub:Primary-and-complementary-indicators} discusses primary and
complementary indicators in further detail. The ``Point of View'' corresponds to
the concept of Evaluation Viewpoint. The ``Context'' facet corresponds to the
SPI Target Entities which set, as defined in
Section~\ref{sub:Gap-analysis-of-evaluation-quality}, the scope of the SPI
implementation.

\begin{example}[ht]
\caption{Determination of measures}\label{exa:Example-5}
Based on the scoping decisions made in Example Box~\ref{exa:Example-4}, ALPHA
needs to identify success indicators (SI) for the Process (PRC) and Project
(PRJ) measurement levels (we omitted the Product level due to space
limitations). Representatives from the three Evaluation Viewpoints (EV) meet to
elicit success indicators for the improvement initiative (code inspections). The
Implementer (I) is represented by one role from the development team and by one
of the project managers; the Coordinator (C) by one representative from the SEPG
and one product manager; and the Sponsor (S) is represented by one member of the
SPI steering committee and the head of the department. The success indicators
can either be primary (p) or complementary (c) from the viewpoint of the
evaluator.\medskip{}

\begin{center}
\begin{tabular}{clcc}
\toprule 
\multicolumn{1}{c}{ML} & \multicolumn{1}{c}{EV} & \multicolumn{1}{c}{SI} &
\multicolumn{1}{c}{MG}\tabularnewline
\midrule
 & Development team (I) & Effectiveness (p) & MG01\tabularnewline
\cmidrule{2-4} 
PRC & SEPG (C) & Effectiveness (p) & MG02\tabularnewline
\cmidrule{3-4} 
 &  & Efficiency (c) & MG03\tabularnewline
\cmidrule{2-4} 
 & SPI steering committee (S) & Effectiveness (p) & MG04\tabularnewline
\midrule
 & Development team (I) & Defects (p) & MG05\tabularnewline
\cmidrule{2-4} 
 & Project manager /  & Defects (p) & MG06\tabularnewline
\cmidrule{3-4} 
PRJ & SEPG (C) & Cost (c) & MG07\tabularnewline
\cmidrule{3-4} 
 &  & Productivity (c) & MG08\tabularnewline
\cmidrule{2-4} 
 & SPI steering committee /  & Defects (p) & MG09\tabularnewline
\cmidrule{3-4} 
 & Head of department (S) & Cost (c) & MG10\tabularnewline
\bottomrule 
\end{tabular}\medskip{}
\par\end{center}

The above table acts as a bridge that pulls together the elements from SPI-MEF
to produce an instance of GQM for the purpose of evaluating an SPI initiative.
The aim of the table is thereby to provide a structure for the forthcoming
derivation of measurement goals (MG). The proposed procedure creates, in
general, a high number of MGs (10 in this example). They can be grouped
according to their respective EVs, that is, MG01 and MG05 belong to the
Development team (I), MG02 and MG07 belong to the SEPG (C), etc. MG01 can be
formulated as: \emph{Evaluate the process of code inspections with respects to
its effectiveness on pre-test defect detection in the pilot project from the
viewpoint of the development team.} The following application of the GQM method
(see for example van Solingen and
Berghout~\cite{van_solingen_goal/question/metric_1999} for detailed
instructions) benefits from the concrete specification of measurement goals.

For the identified measures at the Process level, ALPHA could not define any
baseline as code inspections were introduced for the first time in the pilot
projects (Phase I of the improvement initiative). The collected measures are
however used to establish a baseline against which Phase II is evaluated (see
Example Box~\ref{exa:Example-2} where the initiative context was defined).
Project level measures, on the other hand, are available from two previous
projects. One project manager of the pilot projects and one expert from the SEPG
where nominated as metric evaluators (see Example Box~\ref{exa:Example-8}).
\end{example}

\subsubsection{Primary and complementary
indicators.}\label{sub:Primary-and-complementary-indicators}

\begin{table}
\caption{Identifying complementary
indicators\label{tab:Identifying-complementary-indicators}}

\centering{}
\begin{tabular}{m{1.9in}>{\centering}m{1.1in}>{\centering}m{
0.8in}>{\centering}m{1in}}
\toprule 
 & \emph{Primary} &
\multicolumn{2}{>{\centering}m{1.8in}}{\emph{Complementary}}\tabularnewline
\midrule
Success indicator (Project level) & Cost & Quality & Schedule\tabularnewline
\midrule 
Example metric & Effort in man hours & Defect density & Project cycle
time\tabularnewline
\bottomrule 
\end{tabular}
\end{table}

We define primary indicators as the set of measurements that are used to assess
if the improvement goal has been reached. For example, given that the
improvement goal is cost reduction, primary measurements could be elicited from
the process (e.g. efficiency of the changed/added process) and project level
(e.g. effort in man hours). 

Complementary indicators capture the effects of process improvement that are not
directly connected with the expected effect of the initiative. In other words,
complementary indicators assess the side-effects that may arise when, through an
improvement initiative, the corresponding primary indicator is affected (see
Table~\ref{tab:Identifying-complementary-indicators}). This is needed in order
to control measurement dysfunction that may arise from either wrongly reported
data or from sub-optimization of primary indicators. Iversen and
Mathiassen~\cite{iversen_cultivation_2003} reported on a case where the
measurement program was threatened due to mistrust in the the collected data.

The method we propose to identify complementary indicators borrows its central
idea from the project management
triangle~\cite{atkinson_project_1999,kerzner_project_2009} whose respective
edges represent cost, time and quality. The aim of the project management
triangle is to create the awareness that the entities at the edges are
interrelated with each other and changing one will inevitably affect the others.
Considering this principle in the context of process improvement evaluation,
helps to identify complementary indicators.
Table~\ref{tab:Identifying-complementary-indicators} exemplifies this idea where
cost is a primary success indicator, and quality of the produced artifacts and
project schedule are complementary indicators.

Three basic success indicators, cost, time and quality can be used as a starting
point, since those are the commonly targeted improvement goals, e.g. by Basili
et al.~\cite{basili_sels_1995}, Debou and
Kuntzmann-Combelles~\cite{debou_linking_2000}, Murugappan and
Keeni~\cite{murugappan_quality_2000},
Weiss et al.~\cite{weiss_goal-oriented_2002}, and Moreau et
al.~\cite{moreau_software_2003}. In~\cite{unterkalmsteiner_extended_2011} we
provide an initial set of the success indicators shown in
Table~\ref{tab:Mesurement-Levels-and-success-indicators} that can be refined,
depending on the actual improvement goal(s) and the concrete context in which
the initiative is conducted. 

\begin{table}
\caption{Measurement Levels and success indicators 
(based on~\cite{unterkalmsteiner_evaluation_2012})
\label{tab:Mesurement-Levels-and-success-indicators}}

\centering{}
\begin{tabular}{ll>{\raggedright}p{2.7in}}
\toprule 
Measurement Level & Success Indicator & What is measured?\tabularnewline
\midrule 
Process & Efficiency & The means of the process implementation.\tabularnewline
\cmidrule{2-3} 
 & Effectiveness & The ends of the process implementation, visible in any work
product
and/or artifact.\tabularnewline
\midrule 
Project & Defects & Artifact quality w.r.t. the different phases in the project
life-cycle.\tabularnewline
\cmidrule{2-3} 
 & Cost & Investment in terms of resources and effort to conduct the
implementation
of project.\tabularnewline
\cmidrule{2-3} 
 & Schedule & Calendar time of project and/or phases therein.\tabularnewline
\cmidrule{2-3} 
 & Productivity & Effort input and size output in project
activities.\tabularnewline
\cmidrule{2-3} 
 & Estimation accuracy & Difference between planned and actual outcomes of
project success
indicators.\tabularnewline
\midrule 
Product & Quality & Internal and external quality attributes of the software
product.\tabularnewline
\cmidrule{2-3} 
 & Cost & Total cost of product development \emph{and}
maintenance.\tabularnewline
\cmidrule{2-3} 
 & Time to Market & Calendar time between product inception and
delivery.\tabularnewline
\midrule 
Organization & Economics & Costs and benefits (including intangible
assets)\tabularnewline
\cmidrule{2-3} 
 & Employees & Employee satisfaction \tabularnewline
\cmidrule{2-3} 
 & Growth & Organizational growth, revenue and innovation.\tabularnewline
\cmidrule{2-3} 
 & Communication & Collaboration and communication between employees and/or
customers.\tabularnewline
\midrule 
External & Customer externalities & Any of the above, applied however to the
customers' context. \tabularnewline
\cmidrule{2-3} 
 & Society externalities & Effects on the environment of the
organization.\tabularnewline
\bottomrule
\end{tabular}
\end{table}

\subsubsection{Metrics baselining.}
The success indicators and their respective metrics need to be baselined in
order to serve as the initial point for evaluating the improvement. There are
various ways how organizations can set the baselines for their metrics. The most
typical way would be creating the baseline from historical data collected from
previously conducted processes or projects and already finished products. Some
derived metrics that consist of two or more elementary measurements can
sometimes be easily acquired from historical data. If there is no historical
data available, the baseline can be obtained by collection of data from active
projects that are currently running in the organization. The data collected from
the active projects would serve as the baseline to evaluate the projects that
are going to incorporate the SPI initiative. 

With the definition of the baseline, an expert in interpreting the metric needs
to specify the ranges indicating improvement, stagnation and decline of that
metric. The same evaluator assesses, later on in the evaluation
(Section~\ref{sub:Analysis}), the change of the metrics' value with respect to
the defined baseline.

\subsection{Selection of evaluation
strategies}\label{sub:Selection-of-evaluation-strategies}
\begin{example}[b]
\caption{Evaluation strategy}\label{exa:Example-6}
As the initiative is confined initially to pilot projects, ALPHA decides to
select the basic comparison strategy. In Phase II, when the initiative is rolled
out to all projects within the organization, the survey strategy will be used to
assess the long-term effects on the developed products. As the organization
already conducts customer surveys on a regular basis, they can serve as a
baseline. Before commencing Phase II, the survey instrument is updated to
include the measures defined for the Product level.
\end{example}

SPI evaluation strategies can be classified into four general 
categories~\cite{unterkalmsteiner_evaluation_2012}, that are, in practice, 
often applied in combination: basic comparison, statistics-based analysis, 
survey, and cost-benefit analysis. The fundamental idea of the basic comparison 
strategy is to quantify the impact of an improvement initiative by assessing 
the change of measurements relative to a baseline. In statistics-based 
analysis, statistical tools help to identify and control variation in processes 
over time; surveys are used to collect information on the improvement from 
people who are either directly (employees) or indirectly (customers) affected; 
cost-benefit analysis helps to quantify the financial impact of the SPI 
initiative. 

As illustrated in Figure~\ref{fig:Conceptual-map}, the \emph{Gap analysis of
evaluation quality }concept constrains which evaluation strategies may be
eligible for the specific SPI initiative.
Table~\ref{tab:Criteria-for-selecting-evaluation-strategies} summarizes the
criteria on which the evaluation strategy should be selected. The criterion
``Measurement Levels'' identifies a strategy depending on the selected success
indicators and the corresponding Process, Project, Product, Organization and
External level (see Table~\ref{tab:Mesurement-Levels-and-success-indicators}).
The ``Cost'' criterion provides a relative rank of the required resources to
perform the corresponding evaluation strategy.

\begin{table}
\begin{threeparttable}[b]
\caption{Criteria for selecting evaluation
strategies\label{tab:Criteria-for-selecting-evaluation-strategies}}
\begin{tabular}{llll}
\toprule 
\multicolumn{1}{c}{Strategy} & \multicolumn{1}{c}{Measurement Levels} & Cost &
Confounding factors \\
\midrule
Basic comparison & Process, Project, Product & Medium \tnote{1} & controllable\\
\midrule 
Statistics-based analysis & Process, Project, Product & High & controllable \\
\midrule 
Survey & Product, Organization, External & Low & challenging \\
\midrule 
Cost-Benefit analysis & Product, Organization, External & Medium & challenging
\\
\bottomrule
\end{tabular}
\begin{tablenotes}
 \item [1] if metric collection is not factored in
\end{tablenotes}
\end{threeparttable}
\end{table}

\begin{table}
\caption{Confounding factors (adapted
from~\cite{unterkalmsteiner_evaluation_2012})\label{tab:Confounding-factors}}
\begin{tabular}{>{\raggedright}p{1in}>{\raggedright}p{4.3in}}
\toprule 
Factor & Description\tabularnewline
\midrule 
Project type & New development and enhancement/maintenance projects have
different properties and they should not be treated as the same during
evaluation, i.e. comparison of success indicators from different project types
should be avoided.\tabularnewline
\midrule 
Development model & Different project development life-cycles such as waterfall
model and spiral model have different project characteristics and potentially
confound the evaluation. \tabularnewline
\midrule 
Product size and complexity & Product size (lines of code, function points,
etc.) and complexity (number of features, cyclomatic complexity, etc.) have to
be taken into consideration during the evaluation.\tabularnewline
\midrule 
Product domain & The product domain difference also affects the evaluation.
Front-end applications, server-side systems and embedded software are different
types of product domains that should not be put on par during the evaluation.
\tabularnewline
\midrule 
Technology factors & Technological factors such as the programming language and
tool support can influence indicators like productivity and effort.
\tabularnewline
\midrule 
Process compliance & The degree to which the standard process is followed in the
actual implementation should be considered in the evaluation as this can give
indications to what extent the improvement can be actually attributed to the SPI
initiative. \tabularnewline
\midrule 
Employee factors & The staff working in the project might differ in experience
level and measurement on productivity and efficiency should take staff
experience into consideration. In addition to that, employee turnover in the
organization may also affect the evaluation result. \tabularnewline
\midrule 
Time factors & Time can be seen as a factor that can affect the evaluation
result. When conducting a customer survey on product quality, the time that the
product has been in use needs to be considered.\tabularnewline
\midrule 
Multiple improvement initiatives & It is difficult to ensure that a particular
improvement is attributed to a specific SPI initiative. Several improvement
initiatives that run in parallel would create traceability issues in the
evaluation. For example, when calculating the cost saving from a specific
improvement initiative, care should be taken not to count the saving twice as
the saving might also be attributed to another improvement initiative.
\tabularnewline
\bottomrule
\end{tabular}
\end{table}

The accuracy of the evaluation is influenced by the extent to which the last
criterion, ``Confounding factors'', can be controlled. Confounding factors
represent a fundamental threat for the evaluation of a process improvement
initiative if any kind of comparison is used to assess its effects. A comparison
is said to be confounded if the observed difference between two values (the
effect of a treatment) is not solely caused by the treatment, but can be
partially attributed to an external factor~\cite{pearl_why_1998}.
Table~\ref{tab:Confounding-factors} summarizes typical confounding factors
encountered in the evaluation of improvement initiatives. 

Looking at Table~\ref{tab:Criteria-for-selecting-evaluation-strategies}, we
assess the confounding factors for the ``Basic comparison'' and
``Statistics-based analysis'' strategies as controllable. For example, in the
case of the ``Basic comparison'' strategy, it is common to apply the matching
technique or linear regression models~\cite{unterkalmsteiner_evaluation_2012}.
On the other hand, controlling confounding factors in ``Cost-benefit analysis''
or ``Survey evaluation'' strategies is more challenging. Surveys collect
quantitative and qualitative data from human subjects. Hence it is important to
create a profile of the surveyed individuals in order to group the acquired data
into homogeneous categories. In the cost-benefit analysis strategy, it is
crucial to quantify both direct and indirect costs, and tangible and intangible
benefits~\cite{unterkalmsteiner_evaluation_2012}. 

As we discussed in Section~\ref{sec:Effect-Traceability}, the traceability
of improvement initiatives decreases along the Measurement Levels due to timing
and isolation issues. A viable approach to compensate for the effects of
multiple improvement initiatives is to let internal experts weigh the
contribution of individual initiatives~\cite{van_solingen_measuring_2004}.
Confounding factors related to timing and potential solutions are discussed in
Evaluation implementation (Section~\ref{sub:Evaluation-implementation}).

\subsection{Evaluation implementation}\label{sub:Evaluation-implementation}
The goal of the improvement evaluation is to satisfy the information needs of
the respective stakeholders defined in evaluation scoping (see
Section~\ref{sub:Evaluation-scoping}).

In SPI-MEF, an improvement evaluation is conducted according to a planned
schedule, consisting of the analysis of measures collected at a certain
Measurement Level, and requires the involvement of roles with the expertise to
judge the impact of the improvement initiative. Therefore, each evaluation
instance is assigned a time, a Measurement Level and one or more experts that
conduct the evaluation.

\subsubsection{Scheduling.}\label{sub:Scheduling}
\begin{example}[h]
\caption{Scheduling}\label{exa:Example-7}
ALPHA defined the initial evaluation schedule for 12 months after the
introduction of code inspections. The schedule serves as an indication and may
be updated to the actual progress in the projects.

\emph{ProductA - Feature release 1 - Pilot projects}

The first feature release comprises two projects (Pilot 1 and 2 in Figure (a))
in which code inspections are piloted. The inspections are held monthly on a
sample of the newly implemented code. As the practice requires some training,
the Lag Factor is estimated to 4 months and evaluation $a$, establishing a
baseline for the Process Level measures, is scheduled accordingly. At month 6,
evaluation $b$ is scheduled to assess the Process level again and compare it
with evaluation $a$.

\begin{center}
\includegraphics[scale=0.7]{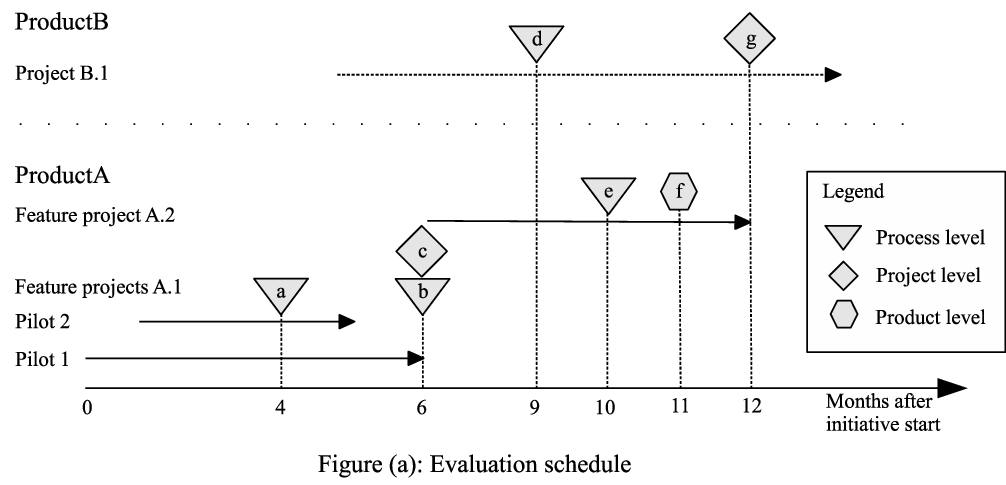}
\par\end{center}

Simultaneously, evaluation $c$ assesses both pilots on the Project level by
comparing the collected measures with the expert estimates from a previous
project in ProductA (no historic data was available and the estimates were
elicited when the measures where defined). The outcome of evaluation $c$ serves
as a baseline on the Project level measures.

\emph{ProductA - Feature release 2}

At month 10 code inspections were introduced in all projects. In evaluation $e$,
as the DF was estimated to 6 months, the baseline of Process Level measures is
updated. At month 11, a survey is initiated to elicit customer satisfaction on
ProductA after the first feature release. Thereby it is assumed that a LF of 5
months (after feature release) is needed to receive reliable feedback (the
specific customers are known to ALPHA). The results of the survey (evaluation
$f$) are added to the Product level baseline.

\emph{Product B}

Development of Product B commenced at month 5 (see Figure (a)). The first
Process level evaluation ($d$) is scheduled for month 9. As evaluation $a$
occurred in month 4, it could still be used as a baseline (DF = 6 months). There
is however evaluation $b$ at month 6 which is then used as baseline for
evaluation $d$. At month 12, the Project level measures are evaluated against
evaluation $c$ (the current baseline), and, as the DF for the Project level was
estimated to 6 months, the result of evaluation $g$ is defined as the new
baseline for the Project level.
\end{example}
The motivation to plan the evaluation schedule is to introduce means to control
timing as a potential confounding factor. The principle idea behind this is that
the effects of a certain improvement initiative may be measurable at different
points in time, depending which Measurement Level is considered in the
evaluation.

The temporal distance in the Measurement Levels (see
Section~\ref{sec:Effect-Traceability}) supports the idea of a lag factor, which
is also referred as the time lag between the cause (SPI initiative) and the
corresponding effect
(improvement)~\cite{norreklit_balance_2000,dror_process_2007}. This latency
needs to be considered when determining the appropriate time to evaluate. 

On the other hand, one has also to consider how long a measurement result is
valid, i.e. how long can it be of value to support decision making processes and
be representative for what is actually assessed\footnote{For example, the
standard CMMI appraisal method for software improvement
(SCAMPI)~\cite{_standard_2011} defines a degradation factor of 3 years for
class A appraisals.}? Due to this validity decay of
measurement results, periodic evaluations are needed in order to make the
effects of the improvement visible over time, as exemplified by Herbsleb et
al.~\cite{herbsleb_benefits_1994}, Jarvinen and van
Solingen~\cite{jarvinen_establishing_1999}, Savioja and
Tukiainen~\cite{savioja_measurement_2007}, Jarvinen et
al.~\cite{jarvinen_integrating_1999}, Iversen and
Ngwenyama~\cite{iversen_problems_2006}, and Moreau et
al.~\cite{moreau_software_2003}. 

In SPI-MEF, we use the terms Lag Factor (LF) and Degradation Factor (DF) to
designate the improvement effect latency and, respectively, the validity decay
of measurement results. DF defines how long an evaluation result may support and
be valid for the decision making process. As a consequence, a periodic
evaluation schedule is required (see Example Box~\ref{exa:Example-7}). LF and DF
are determined by the Measurement Level, the conducted improvement initiative,
the degree to which the changes are actually implemented, and external factors
which may stall progress. Dror~\cite{dror_process_2007} proposes statistical
process control tools and data mining techniques to identify causality links
between improvement action and effect. Historical data from organizations could
therefore be used to create context-sensitive heuristics of improvement timings,
i.e. define the Lag and Degradation Factor based either on collected data and/or
on expert opinion gathered from employees.

\subsubsection{Analysis.}\label{sub:Analysis}
The aim of the analysis is to provide an evaluation to which degree a certain
measurement has changed due to the enacted improvement initiative. To this end,
the expert who was assigned to each measure when they were determined (see
Section~\ref{sub:Determination-of-measures}) rates the change compared to the
baseline. The analysis performed at this stage serves as an intermediate product
that is reused when the holistic view is created, as discussed in
Section~\ref{sub:Holistic-view}.

\begin{example}
\caption{Analysis}\label{exa:Example-8}
\begin{minipage}{0.5\linewidth}
Evaluation $b$ is performed at month 6 to assess the initiative at the Process
level (see Example Box~\ref{exa:Example-7}). The project manager of Pilot 1
analyses the effectiveness and efficiency of the introduced code inspections
(according to the metrics derived in Example Box~\ref{exa:Example-5}). The aim
of evaluation $a$, conducted by the same product manager, was to establish a
baseline and to define a deviation range which demarcates improvement,
stagnation or decline of a metric. Evaluation $b$ shows a significant
improvement in the effectiveness of the process, whereas efficiency remained
stable. The analysis results are eventually communicated to the respective
viewpoints that were defined in the evaluation scope. The figure shows the
results of the evaluation (the process metrics are: phase containment
effectiveness for the coding phase, and defect removal efficiency) and the
targeted viewpoints.
\end{minipage}
\begin{minipage}{0.5\linewidth}
\begin{center}
\includegraphics[scale=0.7]{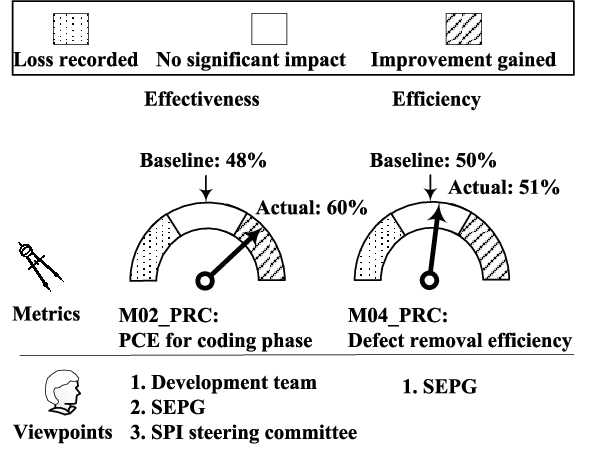}
\par\end{center}
\end{minipage}
\end{example}

\subsection{Holistic view}\label{sub:Holistic-view}
By defining a model which assesses the improvement from the viewpoint of the
involved stakeholders, an important aspect of improvement evaluation can be
addressed, namely increasing the visibility of the improvement initiative as a
whole (\emph{Challenge III - Limited visibility}, see
Figure~\ref{fig:Conceptual-map}\emph{)}. Such a representation would be
beneficial for several reasons. First, the success of an initiative could be
asserted with more confidence since it is assessed considering the involved
stakeholders. Second, it could show, given that the appropriate metrics were
collected, if the improvement has a positive impact on the organization as a
whole or if the change negatively influences aspects which would not have been
considered initially. Third, it can be used as an aid to communicate results of
the improvement in an efficient way, as the amount of data produced in the
individual evaluations is reduced. 

The major aim of the holistic view concept is to provide an aid to communicate
the improvement to the different stakeholders. To achieve this goal, we define
improvement indicators that can be represented in a Kiviat diagram (see Example
Box~\ref{exa:Example-9}). The important information that is shared between the
stakeholders by looking at these diagrams is, that the impact of the improvement
may be different, depending on who is assessing it. Interesting cases are given
if there is a disagreement on the outcome of the initiative, i.e. the evaluation
viewpoints diverge. Such scenarios should give reason for further analysis of
the implemented initiative.

\subsubsection{Considerations for the model.}
In order to show the overall or compound impact of an improvement initiative it
is necessary to define an appropriate model which is able to aggregate the
results of individual evaluations (and metrics) into a representative score. We
identified three basic aspects that need to be considered for such a
construction of the model: 
\begin{enumerate}
\item Normalization of the different metrics to enable a meaningful
aggregation. 
\item Compensation for the different orders of magnitude in the values of the
metrics, i.e. consider that a small difference in one metric may have
effectively more impact than a larger difference in another. 
\item Consideration of the individual viewpoints to include the relative
\textquotedbl{}importance\textquotedbl{} in improving a specific metric. 
\end{enumerate}
The third point has less a technical rather than a qualitative rationale. The
model should take the subjective change, as it was experienced by the involved
parties, into account. This means that each metric should be given a weight,
defined by the viewpoints which are interested in the result of the evaluation.
It is assumed that in this way, the evaluation of the improvement initiative
gains realism by representing the actual situation and reveals possible
imbalances in the change effort, as it was perceived by the involved
stakeholders. 

The first two aspects could be implemented by an impact rating, in which the
evaluator maps the change in a metric into an ordinal scale, which would both
normalize the metrics and compensate the differences in orders of magnitudes.

\subsubsection{Subjective Value of
Improvement.}\label{sub:Subjective-Value-of-improvement}
To calculate the improvement for each Evaluation Viewpoint and Measurement
Level, that is, the Subjective Value of Improvement (SVI), we use two
components. 

The first component is the Subjective Weight (SW) in which each viewpoint
defines a weight of subjective importance to every metric. This means that the
stakeholders of the improvement initiative within the Implementer, Coordinator
and Sponsor viewpoints have to agree on a Subjective Weight. The second
component is the Impact Rating (IR). Here, the expert who conducted the
individual metric evaluations, as presented in Section~\ref{sub:Analysis}, rates
the impact of the improvement initiative on the respective metric according to
an 11 point Likert scale (see Example Box~\ref{exa:Example-9}). One can choose 
also a 7 or 9 point Likert scale, however research suggests that lower than 6 
point scales generally produce less valid scores, have less discriminating 
power and are less preferred by its users~\cite{preston_optimal_2000}. Since 
the Impact Rating is subjective in nature, the organization should discuss and 
agree upon guidelines on how to perform the rating with the aim to improve the 
consistency in the rating between different metric-experts. 

The Subjective Value of Improvement is then calculated as 
\[
SVI=\sum_{id}(SW_{id}*IR_{id})
\]

where $id$ refers to the respective metric identified in the determination of
measures (see Section~\ref{sub:Determination-of-measures}). Since the Impact
Rating component is based on previous evaluations, their Degradation Factor (see
Section~\ref{sub:Scheduling}) determines if the evaluation results are actually
considered at the point in time when the SVI is calculated.

From a measurement theory point of view, the calculation of Subjective Value of
Improvement is questionable as it involves mathematical operations which, in a
strict sense, are not applicable on ordinal scales~\cite{knapp_treating_1990}.
On the other hand, Stevens~\cite{stevens_theory_1946} pointed out that it can be
practical to treat an ordinal scale as an interval scale. Furthermore, several
studies have empirically shown that it matters little if an ordinal scale is
treated as an interval scale~\cite{knapp_treating_1990}. 

\begin{example*}
\caption{Holistic view}\label{exa:Example-9}
ALPHA decided to conduct a holistic evaluation 12 months after code inspections
were introduced as at least one evaluation for each targeted Measurement Level
would have been performed at that time (see Example Box~\ref{exa:Example-7} for
the evaluation schedule). First, the Subjective Value of Improvement (SVI) is
calculated to see how the different Evaluation Viewpoints perceived the
improvement. Then, the Aggregated SVI (ASVI) is used to illustrate the impact of
the initiative on the three Measurement Levels that were scoped for the
evaluation.

The Evaluation Viewpoint assessment uses the SVI as an indicator of the
improvement at a specific Measurement Level. For this example we show the
outcome on the Process level, using the results from evaluation $e$ (see Figure
(a) in Example Box~\ref{exa:Example-7}) as a basis. The Project Manager,
responsible for evaluation $e$, reassesses the impact of code inspections on
feature project A.2, using the Likert scale shown in Figure (b). Representatives
from the involved Evaluation Viewpoints (development team, SEPG, SPI steering
committee) weigh the individual metrics according to their relative importance,
i.e. whether they consider effectiveness or efficiency more critical to fulfill
the aim of the initiative.

\begin{center}
\includegraphics[scale=0.6]{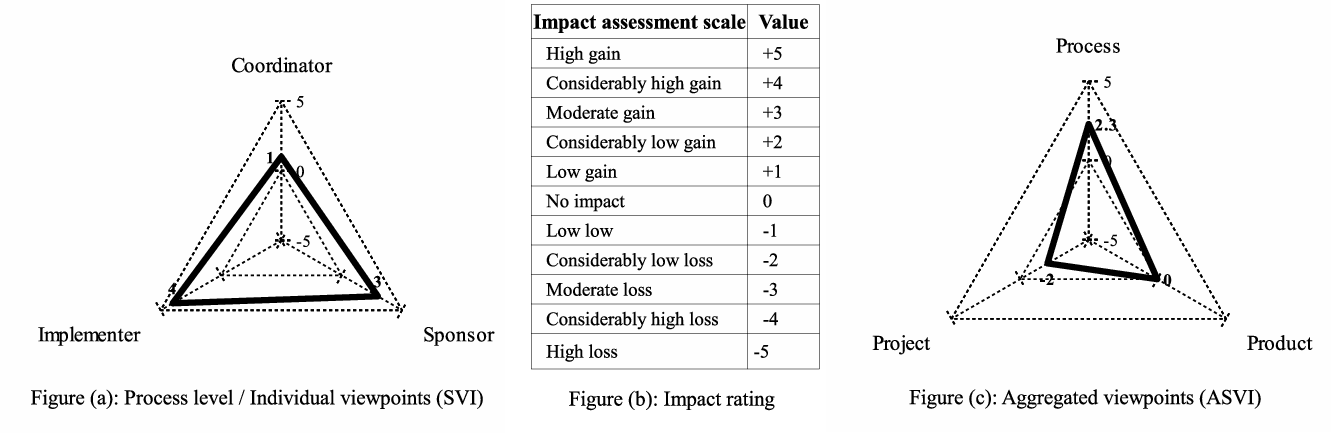}
\par\end{center}

Figure (a) shows that, on the Process level, the Implementer viewpoint perceived
the introduced code inspections more valuable than the Coordinator. Although all
viewpoints identify an improvement, this outcome indicates that the
implementation of the initiative can be further improved. Figure (c) is based on
evaluations $d,e,f,g$ (see Figure (a) in Example Box~\ref{exa:Example-7}), as
those are the only evaluations that are within the DF of 6 months. There, the
Aggregated SVI is calculated on the Process, Project and Product level. Looking
at the Project Level assessment, we can observe a decline in performance. This
indicates that the outcome of the initiative is not coherently seen as a success
by all stakeholders.
\end{example*}

Since the aim of the Holistic View is to provide an overview of the improvement
(the individual evaluations are a more appropriate data source for decision
making), the Subjective Value of Improvement has to be seen as an index or
score. It gives an indication of the improvement, rather than being a metric,
which in its formal definition has to fulfill the representation condition of a
measurement~\cite{fenton_software_1997}. 

\subsubsection{Aggregated Subjective Value of Improvement.}
If evaluations, as presented in Section~\ref{sub:Evaluation-implementation}, are
conducted on different target entities (e.g. projects or products), the
calculation of the Subjective Value of Improvement needs to consider differences
in invested resources. Therefore, the Aggregated Subjective Value of Improvement
(AVSI) is used. The SVI is weighted by an Investment Unit (IU):
\[
ASVI=\sum_{te}(\frac{SVI_{te}*IU_{te}}{IU_{Total}})
\]

where $te$ refers to the Subjective Value of Improvement and the Investment Unit
of the respective target entities (i.e. projects or products). $IU_{total}$ is
the sum of all investments in the target entities. The Investment Unit can be
regarded as the resources which were spent in the implementation of processes,
projects or products on which the individual evaluations are based on. 

\subsection{Summary}
Sections~\ref{sub:Gap-analysis-of-evaluation-quality} to~\ref{sub:Holistic-view}
described the framework, SPI-MEF, targeted at evaluating SPI initiatives.
SPI-MEF aims at providing an SPI evaluation that addresses the challenges
discussed in
Section~\ref{sec:Challenges-in-measuring-and-evaluating-SPI-initiatives}. The
framework is based on several key concepts (Figure~\ref{fig:Conceptual-map})
that span from scoping the evaluation, determining the required measures, to
analyzing the gathered data. In Section~\ref{sec:Validation} we
illustrate how the framework was validated.

\section{Validation}\label{sec:Validation}
The aim of the validation is to determine whether the framework is
able to support practitioners in the evaluation of SPI initiatives.
Section~\ref{sub:Research-method} describes the design of the validation,
whereas Section~\ref{sub:Results} presents its results.
Section~\ref{sub:Threats-to-validity} discusses threats to the validity.

\subsection{Research method}\label{sub:Research-method}
We designed a validation process in which expert judgment from
researchers and industry experts was collected, analyzed and used to refine
SPI-MEF to its final version as it is presented in Section~\ref{sec:SPI-MEF}.
The concrete objectives of this process were:
\begin{enumerate}
\item to identify deficiencies in the proposed concepts
\item to assess the applicability of the framework from a practitioner's point
of view
\item to elicit improvement opportunities for the framework.
\end{enumerate}
As a basis for the validation served a document describing the concepts on which
the framework is based upon.

\subsubsection{Selection of experts.}\label{sub:Selection-of-experts}
\begin{figure}
\begin{centering}
\includegraphics[scale=0.7]{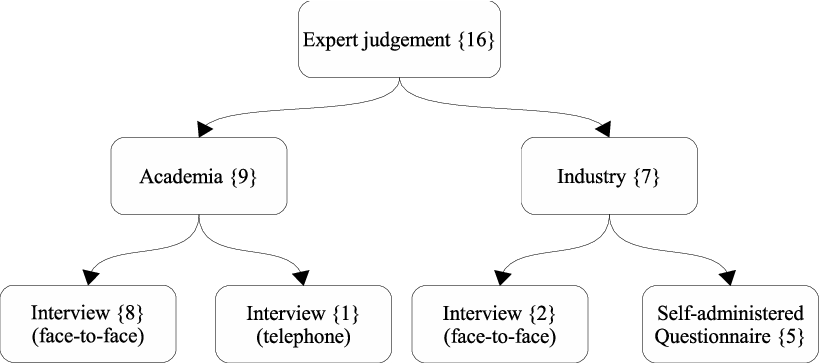}
\par\end{centering}
\caption{Sample of expert judgment and data collection
mechanisms\label{fig:Sample-of-expert-opinion-and-data-collection-mechanisms}}
\end{figure}

We validated the framework by using both researchers and industry experts in
order to address both theoretical aspects and the practicality of the framework.
Since the assessment of the framework's applicability was deemed critical, we
selected researchers with industry experience or working closely with industry.
Figure~\ref{fig:Sample-of-expert-opinion-and-data-collection-mechanisms}
provides an overview how the gathered expert judgment is distributed in
industry and academia, and which data collection mechanisms were employed. The
number in curly braces indicates how many individual experts are included in the
respective category. 

The researchers’ expert judgment was gathered through a semi-structured
interview~\cite{seaman_qualitative_1999}, and was, depending on the
interviewee's accessibility, conducted in a face-to-face meeting or a telephone
call. The group of industry experts was approached either by a semi-structured
interview or a self-administered questionnaire~\cite{jenkins_towards_1997},
again depending on their accessibility.

Thirteen researchers and eleven industry experts were selected and contacted,
whereas nine and seven subjects of the respective groups agreed to participate
in the study. All researchers agreed to provide 45-60 minutes for the interview
and industry experts scheduled a 1½ hour meeting.
Table~\ref{tab:Profile-of-the-industry-and-academia-experts} gives an overview
of the characteristics of the participating subjects.

\begin{table}
\begin{threeparttable}
\caption{Profile of the industry and academia
experts\label{tab:Profile-of-the-industry-and-academia-experts}}
\begin{tabular}{>{\centering}p{1cm}>{\raggedright}p{9cm}>{\centering}p{3cm}}
\multicolumn{3}{>{\centering}p{13cm}}{Academia}\\
\midrule 
\multicolumn{1}{>{\centering}p{1cm}}{Years \tnote{1}} &
\multicolumn{2}{>{\raggedright}p{12cm}}{Research Area}\\
\midrule
8/7 & \multicolumn{2}{>{\raggedright}p{12cm}}{Software Measurement / Estimation,
SPI, Project Management, Requirements Engineering, Business Process Modeling}\\
\midrule 
3/0 & \multicolumn{2}{>{\raggedright}p{12cm}}{Large Scale Software Management,
Software Quality, Product Management, Software Process Management}\\
\midrule 
3/3 & \multicolumn{2}{>{\raggedright}p{12cm}}{Software Product Line Engineering,
SPI, Agile and Lean Software Development, Software Measurement}\\
\midrule 
3/4 & \multicolumn{2}{>{\raggedright}p{12cm}}{Value-Based Software
Engineering}\\
\midrule 
7/3 & \multicolumn{2}{>{\raggedright}p{12cm}}{Software Verification \&
Validation, Search-Based Software Engineering}\\
\midrule 
3/6 & \multicolumn{2}{>{\raggedright}p{12cm}}{Strategic Software Engineering,
Software Product Management}\\
\midrule 
10/0 & \multicolumn{2}{>{\raggedright}p{12cm}}{Requirements Engineering,
Software Architecture}\\
\midrule 
30/10 & \multicolumn{2}{>{\raggedright}p{12cm}}{Software Architecture / Reuse,
Process Engineering and Measurement, SPI}\\
\midrule 
12/8 & \multicolumn{2}{>{\raggedright}p{12cm}}{Verification \& Validation,
Automated Software Engineering, Requirements Engineering, Human / Social aspects
of Software Engineering, Search-Based Software Engineering}\\
\multicolumn{3}{c}{}\\
\multicolumn{3}{c}{Industry}\\
\midrule
Years \tnote{2}
& \multicolumn{1}{>{\raggedright}p{8cm}}{Business Unit} & Company Size
\tnote{3}
\tabularnewline
\midrule
18 & Multimedia & Large\tabularnewline
\midrule
12 & Product Development Excellence & Large\tabularnewline
\midrule
5 & Engineering & Small\tabularnewline
\midrule
5 & Research and Development Engineering & Large \tabularnewline
\midrule
10 & Data Networks Department & Large\tabularnewline
\midrule
12 & Enterprise Mobility Solutions & Large\tabularnewline
\midrule
10 & Research and Development & Large\tabularnewline
\bottomrule
\end{tabular}
\begin{tablenotes}
 \item [1] The values denote the experience in academia/industry
 \item [2] The values denote total work experience
 \item [3] The company size is according to the European Recommendation
2003/361/EC
\end{tablenotes}
\end{threeparttable}
\end{table}

All but one researcher were at the time of the investigation employed at the
Blekinge Institute of Technology; nevertheless, they experienced education in
various universities in Sweden, Germany, Australia and Turkey, which allows the
assumption that their expertise was not streamlined. The industry experts were
employed in four different companies located in Sweden, the U.S., Malaysia and
Singapore. The companies' core businesses were telecommunications, electronic
and electrical manufacturing, and global communication solutions of software
intensive systems.

\subsubsection{Definition of data collection
instruments.}\label{sub:Definition-of-data-collection-instruments}
In this section we describe the design of the interview and the questionnaire.
For the design of the interview instrument we followed the guidelines by
Kvale~\cite{kvale_interviews:_1996}. The interview questions address the
framework's concepts (see Figure~\ref{fig:Conceptual-map}) and were assigned a
priority. The prioritization of the questions is important as the interview
should be designed in a way that it can be finished in the stipulated amount of
time while considering all high priority topics. 

To clarify the purpose of the interview and the contents that need to be
validated, a distilled description (10 pages) of the framework's concepts was
prepared and sent as preparatory documentation to the interviewees. We held also
a short presentation at the beginning of each interview meeting to provide an
introduction to the framework and to give a refresher of its concepts. All
interviews were conducted by the same interviewer to maintain the consistency in
the way the questions are presented. Three note takers recorded the answers
during the interview sessions.

The self-administered questionnaire was designed following Kasunic'
guidelines~\cite{kasunic_designing_2005}. The questions were formulated in a way
such that the respondents could express their degree of agreement/disagreement
(using a Likert scale). Additionally, the respondents were urged to motivate
theirs answers with a few sentences. Furthermore, the wording of the questions
was selected very carefully, taking Salant and Dillman's~\cite{salant_how_1994}
advises into consideration. 

The quality of the interview instrument, the questionnaire and the prepared
supplementary material~\cite{unterkalmsteiner_extended_2011} was improved by
piloting the interview and the questionnaire with three Software Engineering
students. The understandability and clarity in presenting the concepts of the
framework were verified and the questionnaire was assessed regarding question
formulation, layout and the overall compilation process.

\subsection{Threats to validity}\label{sub:Threats-to-validity}
The discussion on the threats to validity of this research is organized
according to the categorization proposed by Wohlin et
al.~\cite{wohlin_experimentation_2000}. Threats to internal validity
(Section~\ref{sub:Interal-validity}) are concerned with the observed
relationship between the treatment and the outcome, i.e. the external factors
that can influence an independent variable with respect to the causal
relationship with a dependent variable. Threats to external validity
(Section~\ref{sub:External-validity}) are factors that can influence the ability
to generalize the results to a wider scope than covered by the study. Construct
validity (Section~\ref{sub:Construct-validity}) is concerned with the
relationship between theory and the observed outcomes of the research, that is,
with the ability to generalize its results to the theoretical construct which
motivate the research. Threats to the conclusion validity
(Section~\ref{sub:Conclusion-validity}) are concerned with factors that affect
the ability to draw the correct conclusions from the conducted study.

\subsubsection{Internal validity.}\label{sub:Interal-validity}
Three threats to internal validity, related to the gathering of expert
judgment, were identified: instrumentation, maturation, and selection. 

The instrumentation threat is caused by bad design of
artifacts~\cite{wohlin_experimentation_2000} used in the expert judgment
elicitation.
Those can lead to misunderstandings regarding the discussed topic and weaken the
results from the gathered data. To minimize this threat, the preparatory
document and the interview questions were piloted first with three Software
Engineering students to test whether the artifacts are clear and understandable.
Afterwards, the preparatory document and the interview questions were refined. 

The maturation threat exists if the experts’ behavior changes during the
elicitation process as the time passes~\cite{wohlin_experimentation_2000}. This
can distort the gathered results if the subjects acquire new knowledge during
the process, or become detached~\cite{wohlin_experimentation_2000}. This threat
is regarded as minor since the interviews with researchers were conducted during
a meeting which lasted approximately one hour each. The written questionnaire
was compiled and returned by all industry experts within two weeks; since no
deadline to return the questionnaire was given to the subjects, the rather quick
response indicates that they were committed to the task and had interest in
providing useful information. Furthermore, the questionnaire was designed to
present the needed information and the questions concisely and precisely such
that it can be compiled within approximately one hour. 

The selection threat is concerned with the varying human performance and
potential biases introduced by the selected subjects for the investigation, e.g.
higher motivation of volunteers may lead to better
results~\cite{wohlin_experimentation_2000}. As the presented profiles in
Section~\ref{sub:Selection-of-experts} show, both researchers and industry
experts have several years of experience in their respective fields. Obviously
there are differences in expertise in the specific areas of interest but this
was regarded rather as an advantage than a threat since a major goal of the
validation was to identify new, not yet considered, aspects for the measurement
and evaluation of SPI. 

\subsubsection{External validity.}\label{sub:External-validity}
The threat of selection and treatment is caused by not having a representative
sample of the population~\cite{wohlin_experimentation_2000}. To address this
threat, the selection of researchers took also the industrial experience of the
subjects into consideration. The industry experts selected in this study were
employed in different companies with different core businesses from Europe, the
United States and Asia. Nevertheless, there is a moderate threat of selection
bias due to the convenience sampling of researchers and industry experts.

\subsubsection{Construct validity.}\label{sub:Construct-validity}
In this category, two threats for this research were identified: mono-operation
bias and evaluation apprehension.

Mono-operation bias is caused by considering only a single subject, independent
variable or case and hence, the study may not fully represent the investigated
theory~\cite{wohlin_experimentation_2000}. This threat is considered moderate
since two groups with different background were considered and for each group
(academic and industry experts) more than one subject was involved. 

The threat of evaluation apprehension is caused by the human tendency to behave
differently while being evaluated~\cite{wohlin_experimentation_2000}. This can
distort the result of the study since the subjects may perform better than in a
regular, unobserved, situation. To tackle this issue, the experts were
guaranteed their anonymity and that their answers were only used by the
researchers involved in the study.

\subsubsection{Conclusion validity.}\label{sub:Conclusion-validity}
Three threats were identified in this study that fall under this category:
random heterogeneity of subjects, random irrelevancies in experimental setting,
and searching for a certain result. 

Random heterogeneity of subjects is a threat caused by a heterogeneous sample
such that individual differences within the sample could affect the study’s
result~\cite{wohlin_experimentation_2000}. To minimize this threat, the experts
were selected based on their competencies and knowledge in software engineering
and software process improvement. 

Random irrelevancies are elements outside of the study setting which can disturb
its conduct~\cite{wohlin_experimentation_2000}. This threat is considered as
minor since the interviews were conducted in an uninterrupted session and in a
quiet environment. There were no discussions about the questions before the
interview that could have influenced the interviewee’s answers. 

Searching for results or “fishing” is the tendency of the researchers to search
for a certain result or answer and ignore the inconvenient
information~\cite{wohlin_experimentation_2000}. To minimize this threat, all
answers from the experts, whether they were positive or negative, were recorded
and analyzed regardless the researchers’ expected outcome. 

\subsection{Results}\label{sub:Results}
The subsections~\ref{sub:validation-Gap-analysis-of-evaluation-quality}
to~\ref{sub:validation-Holistic-View} summarize the main issues
regarding the presented concepts (see Figure~\ref{fig:Conceptual-map}). The
impact on SPI-MEF and the applied refinements are reported in each subsection.

In essence, the approach proposed by SPI-MEF for SPI evaluation was taken very
positively by both academic researchers and industry practitioners. Both groups
agreed that SPI-MEF has the potential to provide a systematic way of evaluating
process improvement impact. However, there were several suggestions brought
forward to improve the framework in terms of increasing its applicability in
practice. 

\subsubsection{Gap analysis of evaluation
quality.}\label{sub:validation-Gap-analysis-of-evaluation-quality}
Higher accuracy and better coverage (see
Section~\ref{sub:Gap-analysis-of-evaluation-quality}) is of course good to
achieve. However, it may not be feasible for companies to achieve both
simultaneously in the first place since resources may be constrained. Therefore,
it is important to know which one is important to consider first. There was
divergence in the answers of the interviewees on this issue. Some suggested
considering accuracy first while others considered coverage as more important.
However, their answers revealed that giving emphasis on accuracy first has some
formidable advantages.
Achieving accurate and valid results first can increase the confidence on the
quality of evaluation which then can motivate to increase the coverage adding
more complexity in the evaluation and investing more resources. If the intention
of the evaluation is to see a more complete picture of the improvement benefits
first and identifying the problem areas, then coverage should get more emphasis
than accuracy.

The cost of the evaluation was considered as a very important factor. The
absence of cost considerations may lead organizations to opt for a good enough
evaluation and discourage them from expending money to gain high accuracy and
coverage to achieve a holistic evaluation. Therefore, the cost factor should be
included in this matrix and a discussion on the relation between quality of
evaluation and cost should be included in the concept.

\emph{Impact on SPI-MEF:} In addition to the previously present two dimensions
of accuracy and coverage, a third dimension covering the cost aspect was added
to the framework (see Section~\ref{sub:Gap-analysis-of-evaluation-quality}).

\subsubsection{Evaluation scoping}
Some confusion arose regarding the roles in each viewpoint and in the
interpretation of the categorization of the viewpoints (see
Section~\ref{sub:Evaluation-viewpoints}). For example, it was not clear that the
same role can subsume different viewpoints (e.g. a project manager who has the
viewpoint of a Coordinator in the Project level can also have the Implementer
viewpoint in the Product level). This may be due to the short document provided
to the experts as an introduction to the framework which did not suffice to
clarify this aspect.

\emph{Impact on SPI-MEF:} To reduce chances of misinterpretation, the example
describing the Evaluation Viewpoints in Section~\ref{sub:Evaluation-viewpoints}
explicitly discusses this point. The extended
scenario~\cite{unterkalmsteiner_extended_2011} was enhanced with motivations for
the allocation of roles to the different viewpoints.

\subsubsection{Determination of measures}
The feedback to the questions regarding the proposed method (see
Section~\ref{sub:Determination-of-measures}) was twofold: on one hand, the
approach was judged as systematic and comprehensive, which indicates that the
method can be of practical use and provide appropriate support for
practitioners. On the other hand, some experts perceived the approach as quite
complex and time consuming which implies caveats in its applicability in terms
of training and education of employees, and in justifying the additional
resources needed for its implementation. 

\emph{Impact on SPI-MEF:} The concerns about the complexity of the method can be
addressed by considering that the process of derivation of measures is an
iterative one and is indeed scalable to more realistic settings than those which
were shown in the example given in the interview material. Adding to the
framework, as it was proposed by one interviewee, a palette of goals, questions
and metrics on which the user can base his measurement program, was regarded by
the authors as inflexible and difficult to maintain. It would be more
appropriate to define a step-by-step guide which leads the practitioner to
formulate his own goals and questions and then provide a pool from which he can
pick the needed metrics (see
Section~\ref{sub:Primary-and-complementary-indicators}). Clearly, this implies
more effort on the part of the user of the framework; however, this approach
makes it flexible and applicable in a wider range of scenarios.

\subsubsection{Primary and complementary measures}
The introduction of \textquotedbl{}primary\textquotedbl{} and
\textquotedbl{}complementary\textquotedbl{} measurements (see
Section~\ref{sub:Primary-and-complementary-indicators}) necessitates a precise
definition of these new terms. As it was observed by one academic expert, the
term \textquotedbl{}complementary\textquotedbl{} may induce misunderstandings,
and indeed, an industry expert interpreted the measures as the
\textquotedbl{}needed\textquotedbl{} (primary) and \textquotedbl{}good to
have\textquotedbl{} (complementary) ones. Clearly, this was not the intended
interpretation and several remedies were discussed to avoid this
misinterpretation.

\emph{Impact on SPI-MEF:} As a result, a renaming of the terms was discarded,
since any naming inherits ambiguities depending on the background of the reader.
Therefore, in order to minimize the space for interpretation, the definition of
the terms \textquotedbl{}primary\textquotedbl{} and
\textquotedbl{}complementary\textquotedbl{} were enhanced and the exemplified
measurement derivation was elaborated with more detailed steps. Additionally, it
was made very explicit in the framework that
\textquotedbl{}complementary\textquotedbl{} measures are not optional
(\textquotedbl{}good to have\textquotedbl{}), but necessary for a complete
evaluation (Section~\ref{sub:Primary-and-complementary-indicators}). 

Furthermore, a pool of commonly used metrics, grouped according to measurement
levels, was provided in~\cite{unterkalmsteiner_extended_2011}. This should
support the practitioner initially in identifying primary and complementary
measurements. It should be noted however, that the pool has to be seen as a
reference, and it should not be regarded as an exhaustive set of metrics. 

\subsubsection{Confounding factors}
This concept (see Section~\ref{sub:Selection-of-evaluation-strategies}) was
specifically put to the industry experts in order to exhibit if they consider it
as an important issue in the practical evaluation of SPI. Compared to the other
questions, the input to this concept was rather thin, although positive. Indeed,
it was deemed as a necessary step to create awareness for confounding factors
and consider them appropriately in the construction of baselines, and practical
ways to control them in an industrial setting were needed according to the
industry experts.

\emph{Impact on SPI-MEF:} In the final framework, a short description of typical
confounding factors (Table~\ref{tab:Confounding-factors}) that need to be taken
into consideration for evaluation planning or during the evaluation was
included, along with guidelines on how to address them
(Section~\ref{sub:Selection-of-evaluation-strategies}).

\subsubsection{Evaluation scheduling}
Both the academic and industry experts agreed that the concept of Lag Factor
(LF) and Degradation Factor (DF), and periodic evaluation is conceptually right
(see Section~\ref{sub:Scheduling}). The main concerns were, however, how to come
up with these values in the first place when the initiative is new or when
several improvement initiatives are running in parallel. DF was considered
harder to define as compared to LF. DF is the key concept that helps to define
the time-bounds for periodic evaluation and could also help to determine the
optimum interval between successive evaluations which is important to minimize
the cost of evaluation.

\emph{Impact on SPI-MEF:} It was suggested to provide some guidelines on how to
come up with the values of LF and DF. These could however be misleading as long
as empirical evidence or heuristics for LF and DF are not available. Therefore,
at the beginning when the framework is introduced in an organization,
experienced practitioners and experts in the field of process improvement could
help to define these values. Thereafter, organizations can learn and improve
their accuracy to determine LF and DF when they gain more and more empirical
evidence for appropriate values of LF and DF.

\subsubsection{Holistic view}\label{sub:validation-Holistic-View}
The scrutiny of the Holistic view concept (see Section~\ref{sub:Holistic-view})
revealed some important characteristics regarding this approach to present
improvement, and which strengths and weaknesses are inherent in this approach.
It was confirmed that the target audience for the holistic representation
resides in top-level management, for which the reduction in details can be seen
as an advantage. The tool is therefore less adequate as decision support for the
continuation or further refinement of an improvement initiative (this has to be
done at a lower level where details are conserved), but rather expresses the
\textquotedbl{}health\textquotedbl{} on the initiative and reveals if the
expected benefits are achieved. The subjective element, \textquotedbl{}gut
feeling\textquotedbl{}, as it is integrated in the model, was judged both
positively and negatively. Subjective ratings in improvement assessment are used
in industry and therefore applicable in the \textquotedbl{}Holistic
View\textquotedbl{}. The contribution of the framework would therefore be the
formalization of that process.

\emph{Impact on SPI-MEF:} To make the subjective rating in the improvement
assessment more homogenized and consistent among the different stakeholders, the
framework prescribes to create guidelines on how to perform such a rating
(Section~\ref{sub:Subjective-Value-of-improvement}). The extended
scenario~\cite{unterkalmsteiner_extended_2011} provides an example how such a
guideline can be realized as a help to homogenize impact rating.

\section{Conclusion}\label{sec:Conclusion}
This paper presents a framework for the measurement and evaluation of software
process improvement initiatives (SPI-MEF). SPI-MEF describes and exemplifies the
use of the concepts of evaluation quality and scoping, determination of
measures, and evaluation scheduling and analysis. The framework's concepts were
derived from the best practices gathered in a systematic literature review on
SPI measurement and evaluation~\cite{unterkalmsteiner_evaluation_2012}. 

Once the framework was created initially, it was evaluated by sixteen academic
and industry experts with a median of 6 years of combined SPI experience in both
research and practice. The focus of the evaluation was to validate that the
framework integrates the important aspects of SPI evaluation and, on the other
hand, provides support for practitioners. According to the experts, the
contribution of the framework lies in the structured and nevertheless flexible
approach. 

SPI-MEF gives concrete guidelines on how to scope the evaluation \emph{before}
the improvement initiative is implemented. This allows practitioners to increase
the visibility of the improvement effort within the company and to plan the
required resources needed for the evaluation. As SPI-MEF builds upon the widely
known GQM paradigm, existing measurement programs in an organization can be
re-focused on the evaluation of SPI, and hence reusing existing resources and
infrastructure. On the other hand, SPI-MEF provides also guidance to initiate a
new measurement and evaluation program.

Perception of improvement success varies within the functional structure of an
organization. SPI-MEF provides means to capture and communicate improvement
outcomes from different viewpoints, facilitating the understanding of the
effects of process change. As such, SPI-MEF is a step forward in the ability to
determine the success of SPI initiatives, increasing the confidence in the
evaluation results.

\subsection{Future work}
Future work, refining and extending SPI-MEF, will include the integration of a
cost-model that will further increase the adaptability of the framework, and
improve the support for practitioners selecting evaluation strategies.
Furthermore, we target a dynamic evaluation~\cite{gorschek_model_2006} of
SPI-MEF, instantiating the framework in a specific company context and piloting
the implementation within an initiative that aims at improving the alignment
between requirements engineering and verification activities.

\bibliographystyle{wileyj}
\bibliography{references}

\end{document}